\documentclass[12pt,nonbibnotes,nofootinbib,notitlepage]{revtex4-2}
\usepackage{amsmath}
\usepackage{amssymb}
\usepackage{bm}
\usepackage{epsfig}
\usepackage{graphicx}
\usepackage[dvipsnames]{xcolor}
\usepackage[unicode=true,colorlinks=true,citecolor=blue,urlcolor=blue]{hyperref}
\usepackage[T2A]{fontenc}
\usepackage[cp1251]{inputenc}
\usepackage[english]{babel}

\usepackage[normalem]{ulem}
\renewcommand{\sout}[1]{}

\newcommand{\Tr}{\mathop{\rm Tr}\nolimits}

\renewcommand{\Re}{\mathop{\rm Re}}
\renewcommand{\Im}{\mathop{\rm Im}}
\newcommand{\e}{{\rm e}}

\let\ifr\i
\renewcommand{\i}{{\rm i}}
\renewcommand{\d}{\mathrm d}
\newcommand{\braket}[1]{\left\langle #1 \right\rangle}
\newcommand{\erfc}{\mathop{\rm erfc}\nolimits}

\usepackage{color}

\begin{document}


\title{Theory of optically detected spin noise in nanosystems\footnote{This is the translation of the original manuscript in Russian available in the supplementary files at arXiv and at \href{https://www.dropbox.com/s/kz5uyz3btgocf8l/UFN_spin_noise_review_rus.pdf?dl=1}{this URL}.}}

\author{D.~S.~Smirnov$^1$\footnote{smirnov@mail.ioffe.ru}, V.~N.~Mantsevich$^2$, M.~M.~Glazov$^{1,3}$ }
\affiliation{$^1$Ioffe Institute, 194021, Saint-Petersburg \\
$^2$Lomonosov Moscow State University, 119991, Moscow \\
$^3$Spin Optics Laboratory, St. Petersburg University, 198504, Saint-Petersburg
}


\begin{abstract}

Theory of spin noise in low dimensional systems and bulk semiconductors is reviewed. Spin noise is usually detected by optical means, continuously measuring the rotation angle of the polarization plane of the probe beam passing through the sample. Spin noise spectra yield rich information about the spin properties of the system including, for example, $g$-factors of the charge carriers, spin relaxation times, parameters of the hyperfine interaction, spin-orbit interaction constants, frequencies and widths of the optical resonances. The review describes basic models of spin noise, methods of its theoretical description, and their relation with the experimental results. We also discuss the relation between the spin noise spectroscopy, the strong and weak quantum measurements and the spin flip Raman scattering, and analyze similar effects including manifestations of the charge, current and valley polarization fluctuations in the optical response. Possible directions for further development of the spin noise spectroscopy are outlined.


\end{abstract}

\maketitle
\newpage
\tableofcontents
\newpage


\section{Introduction}
Studies of spin phenomena have formed a broad and rapidly developing branch of the solid state physics since the beginning of XXIst century. Such a violent progress is related to, on the one hand, the prospects of realizing quantum methods for information processing and, on the other one, the novel fundamental physics of spin phenomena which manifest themselves in the condensed matter.
Usually, the dynamics of electron and nuclear spin ensembles in semiconductors is studied via monitoring the response of the system to external perturbations, mainly, to alternating electric and magnetic fields. One can detect, using the polarization of emitted light and spin-Faraday and Kerr effects, i.e., the rotation of the probe beam polarization plane at its transmission through the medium or its reflection from it, the spin precession in external fields and spin relaxation resulting from the excitation of the spin polarization, e.g., by circularly polarized light.

\begin{figure}[t]
\includegraphics[width=0.65\linewidth]{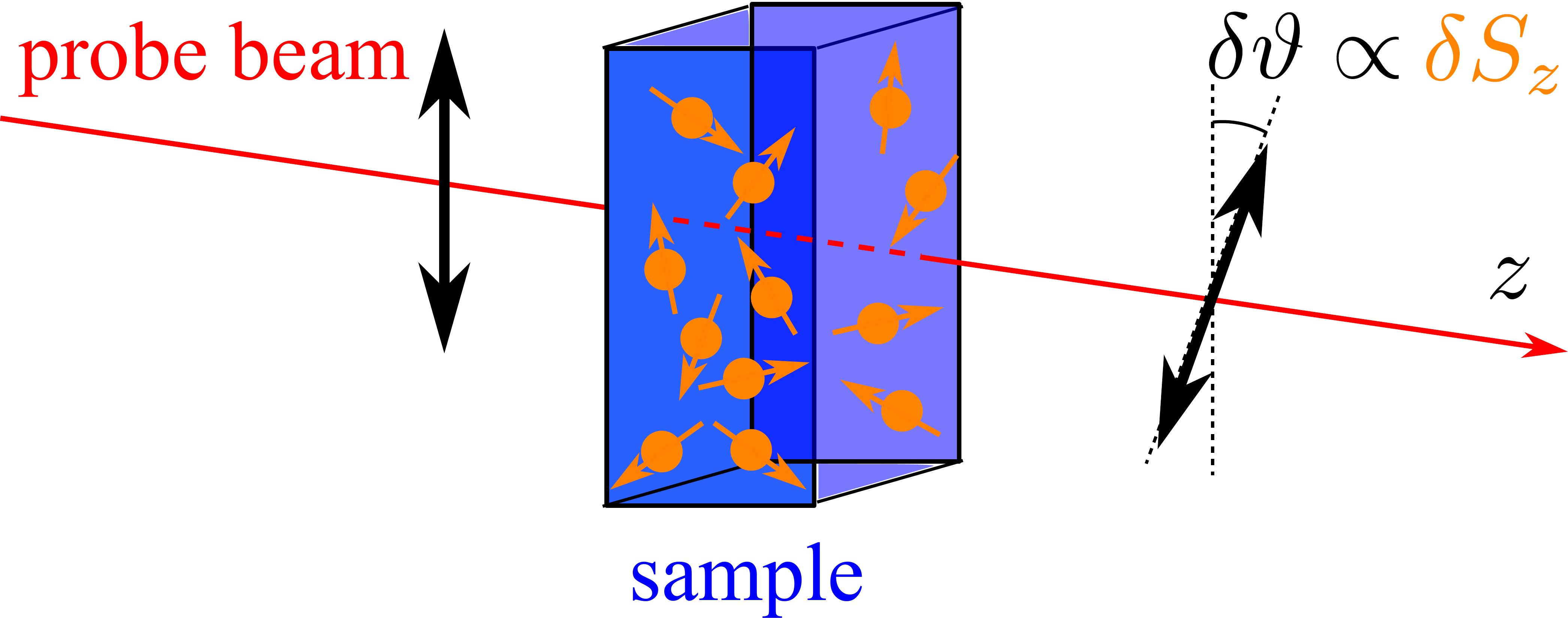}
\caption{Illustration of spin fluctuations detection: $z$ is the direction of linearly polarized light propagation, double arrows show orientation of the linear polarization plane of the beam before and after its transmission through the medium with fluctuating spins. (Adapted from Ref.~\cite{glazov_keldysh}.)
}\label{fig:ch9:SNS}
\end{figure}

It is natural to ask a question about what one can learn about the spin dynamics from studies of the Faraday or Kerr rotation without any excitation of the system. Imminent fluctuations of electron spins, $\delta S_z(t)$, see Fig.~\ref{fig:ch9:SNS}, result in stochastic contribution to the Faraday rotation angle:
\begin{equation}
\label{dtheta:gen}
\delta \vartheta(t) \propto \delta S_z(t).
\end{equation}
Here $z$ is the light propagation axis, and we assume that the retardation effects at the light transmission can be neglected.  By definition, the spin fluctuations are absent on average, $\langle \delta S_\alpha(t) \rangle=0$.  Here the angular brackets denote the time averaging, namely,
\begin{equation}
\label{time:aver}
\langle \delta S_\alpha(t) \rangle = \lim_{T\to \infty} \frac{1}{2T} \int_{-T}^T \delta S_\alpha(t) dt, \quad \alpha=x,~y,~z,
\end{equation}
where $T$ is a macroscopic time which exceeds by far the spin precession period in external fields, times of spin decoherence and relaxation, etc. Spin fluctuations -- spin noise -- are characterized by correlation functions. Of prime interest are the  second order correlation functions under the steady-state conditions~\cite{glazov2018electron}, which are defined as follows:
\begin{equation}
\label{correlations}
\mathcal C_{\alpha\beta}(\tau) = \langle \delta S_{\alpha} (t+\tau) \delta S_{\beta} (t)\rangle.
\end{equation}
In accordance with the general theory of fluctuations~\cite{ll5,ll10} the averaging in Eq.~\eqref{correlations} takes place over the time  $t$ at a fixed difference of the arguments  $\tau$. It is assumed in Eq.~\eqref{correlations} that the quantity $\delta S_\alpha(t)$ is a classical one. Generally, one has to symmetrize quantum mechanical operators~\cite{ll5}
\begin{equation}
\label{correlations:qnt}
\mathcal C_{\alpha\beta}(\tau) = \langle \{\delta \hat S_{\alpha} (t+\tau) \delta \hat S_{\beta} (t)\}_s\rangle,
\end{equation}
where $\{\hat A \hat B\}_s = (\hat A \hat B+\hat B \hat A)/2$, and the spin fluctuation operator $\delta \hat S_\alpha = \hat S_\alpha - \langle \hat S_\alpha\rangle$, with the angular brackets here denoting the quantum mechanical averaging.

Owing to Eq.~\eqref{dtheta:gen}, the autocorrelation function of the Faraday rotation angles $\langle \delta \vartheta_F(t+\tau) \vartheta_F(t)\rangle$ is directly proportional to the spin autocorrelation function  $\mathcal C_{zz}(\tau)$. Thus, observing the fluctuations of the Faraday and Kerr rotation (as well as those of the ellipticity) one can directly measure the spin correlation function. The latter contains fundamental information on spin dynamics. In thermal equilibrium conditions, the functions $\mathcal C_{\alpha\beta}(\tau)$ are related, via the fluctuation-dissipation theorem, to the spin susceptibility of the system. E. B. Aleksandrov and V. S. Zapasskii were first to detect the spin fluctuations and, correspondingly, the spin susceptibility in the noise of Faraday rotation in atomic gas vapors~\cite{aleksandrov81}. These experiments of 1980s were reproduced in the beginning of 2000s~\cite{PhysRevLett.80.3487,Mitsui:2000nx,Crooker_Noise}. Later on, this method has been applied to bulk semiconductors~\cite{Oestreich_noise,PhysRevB.79.035208,Romer2010} and semiconductor nanosystems~\cite{muller-Wells,crooker2010,PhysRevLett.108.186603,singleHole}. As a result, a novel area of spin dynamics studies, known as spin noise spectroscopy, has been formed~\cite{dyakonov_book,1742-6596-324-1-012002,Mueller2010,Zapasskii:13,Oestreich:rev,sinitsynreview}.

For spin fluctuations detection, superconducting interferometric devices~\cite{PhysRevLett.55.1742,PhysRevLett.57.905}, scanning tunneling microscopy~\cite{Manassen,doi:10.1063/1.1434301}, and magnetic resonance microscopy~\cite{rugar2004single,PhysRevLett.99.250601} can be applied. However, in semiconductor systems optical methods described above are the most efficient.

In the next section of the review a consistent description of the simplest methods for calculating the spin fluctuations will be presented using, as an example, the spin precession in an external static magnetic field and isotropic spin relaxation, which is typical for bulk semiconductors. Next, in Secs.~\ref{sec:0D}, \ref{sec:1D}, and~\ref{sec:2D} the specifics of the spin noise in zero-, \mbox{one-,} and two-dimensional systems will be detailed, respectively. In Sec.~\ref{sec:high} high order spin correlation functions will be described separately. Further, in Sec.~\ref{sec:extended} the abilities of the spin noise spectroscopy technique will be described, that go beyond measurements of spin correlators. The concluding Sec.~\ref{sec:concl} contains the summary and description of prospects of further development of the spin fluctuations theory in semiconductors.


\section{Methods of the spin noise spectra calculations}
\label{sec:methods}

\subsection{Methods of the fluctuations theory}

Let us consider fluctuations of a single electron spin $\delta {\bm s}(t)$ in a static magnetic field $\bm B$, which is characterized by a Larmor frequency $\bm \Omega = g\mu_B \bm B/\hbar$, where $g$ is the $g$-factor and  $\mu_B$ is the Bohr magneton. Spin correlation functions can be calculated by a number of methods. We present, at first, the Langevin random forces approach based on the solution of the Bloch equation for spin fluctuation
\begin{equation} \label{ch9:field}
\frac{\partial \delta {\bm s}(t)}{\partial t} + \frac{ \delta {\bm s}(t)}{\tau_s}  + \delta{\bm s}(t) \times {\bm \Omega} = {\bm \xi}(t)\:,
\end{equation}
where $\tau_s$ is a phenomenological spin relaxation time, ${\bm \xi}(t)$ are the random or Langevin forces. These fictitious forces are included in the right hand side of Eq.~\eqref{ch9:field} to support mean square fluctuations of the spin components
\begin{equation}
\label{fluct:single:0}
\langle \delta s_\alpha (t) \delta s_\beta (t) \rangle  \equiv  \langle \{\delta \hat s_\alpha (t) \delta\hat s_\beta (t)\}_s \rangle = \frac{\delta_{\alpha\beta}}{4}.
\end{equation}
This expression is given in the high-temperature limit, $k_B T \gg \hbar\Omega$, where magnetic field induced equilibrium polarization in negligible. Equation~\eqref{fluct:single:0} directly follows from the spin operators definition. Since the Langevin forces are fictitious, i.e., they are not directly related to any real physical process, their correlation function must not contain any temporal scales, i.e., they reduce to the  $\delta$-function~\cite{ll5,Lax1,Lax2,springerlink:10.1007}:
\begin{equation} \label{correlator_force}
\langle \xi_{\alpha}(t) \xi_{\beta}(t+\tau) \rangle = \frac{1}{2 \tau_s} \delta_{\alpha \beta} \delta(\tau)\:.
\end{equation}
Solving Eq.~\eqref{ch9:field} one is able to calculate the spin correlators both in the temporal and frequency representations. By definition,
\begin{equation}
  \label{correlator_spin}
  \tilde{\mathcal C}_{\alpha\beta}(\omega) = (\delta s_{\alpha} \delta s_{\beta})_{\omega} = \int\limits_{- \infty}^{+ \infty} \langle \delta s_{\alpha}(t+ \tau) \delta s_{\beta} (t) \rangle\e^{\i\omega\tau}\d\tau.
\end{equation}
In the coordinate frame $x_1,y_1,z_1$ with the axis $z_1 \parallel {\bm \Omega}$ we have~\cite{NoiseGlazov,glazov2018electron}
\begin{subequations}
\label{noise:field}
\begin{eqnarray}
&&\label{eq:longitudinal}(\delta s_{z_1}^2 )_{\omega} = \frac{\pi}{2} \Delta(\omega)\:, \\
&&\label{eq:transverse}(\delta s_{x_1}^2 )_{\omega} = (\delta s_{y_1}^2 )_{\omega} = \frac{\pi}{4} [\Delta(\omega - \Omega) +
\Delta(\omega + \Omega)]\:,  \\
&&(\delta s_{y_1} \delta s_{x_1})_{\omega} = (\delta s_{x_1} \delta s_{y_1})^*_{\omega} = \frac{2 {\rm i} \omega \Omega \tau_s^2}{1 + \tau^2_s (\omega^2
+ \Omega^2)}(\delta s_{x_1}^2 )_{\omega}\:,
\end{eqnarray}
\end{subequations}
where a broadened $\delta$-function is introduced in accordance with
\begin{equation}
  \label{eq:Delta_def}
  \Delta(x) = \frac{1}{\pi} \frac{\tau_s}{ 1 + (x \tau_s)^2}\:.
\end{equation}
It follows from Eqs.~\eqref{noise:field} that the parallel to the magnetic field spin component is unaffected by the field (the $\bm B$-dependence can arise from the field dependence of the spin relaxation time $\tau_s(B)$). The perpendicular to the field components precess at the frequency $\Omega$. It results both in the shift of the arguments of the broadened $\delta$-functions in $(\delta s_{x_1}^2 )_{\omega}$, $(\delta s_{x_1}^2 )_{\omega}$, and in the appearance of the cross-correlators $(\delta s_{y_1} \delta s_{x_1})_{\omega} = (\delta s_{x_1} \delta s_{y_1})^*_{\omega}$.

The Langevin random forces method is not the only one to calculate spin correlators. In many cases, it is convenient to write and solve kinetic equations for the correlation functions $ \mathcal C_{\alpha\beta}(\tau)$~\cite{Glazov2013:rev}. In the simplest case under consideration, these equations have the form (at $\tau>0$)
\begin{equation}
\label{eq:corr:gen}
\frac{\partial}{\partial \tau} \mathcal C_{\alpha\beta}(\tau) + \sum_{\gamma\delta} \epsilon_{\alpha\gamma\delta} \mathcal C_{\gamma\beta}(\tau) \Omega_{\delta} + \frac{\mathcal C_{\alpha\beta}(\tau) }{\tau_s} =0,
\end{equation}
where $\epsilon_{\alpha\gamma\delta}$ is the Levi-Civita symbol. In agreement with the general approach, Eqs.~\eqref{eq:corr:gen} should be supplemented by the initial condition~\eqref{fluct:single:0}, that describes the single-time correlators. This approach turns out to be particularly convenient for description of spin noise in non-equilibrium conditions~\cite{PhysRevB.90.085303,glazov_keldysh}.

As we already mentioned, the spin correlation functions $\tilde{\mathcal C}_{\alpha\beta}(\omega)$ at thermal equilibrium conditions can be expressed via corresponding components of the spin susceptibility. This statement known as the fluctuation-dissipation theorem can be formulated as follows. Let us introduce generalized forces  $\bm f$ with the Cartesian components
\[
f_\alpha(t) = f_{\alpha,\omega} e^{-\mathrm i \omega t} + c.c.,
\]
which provide a perturbation to the spin system Hamiltonian in the form
\begin{equation}
\label{pert}
\hat V = - \sum_\alpha \hat{s}_\alpha f_\alpha\:.
\end{equation}
We also introduce the spin susceptibility,  $\mu_{\alpha\beta}(\omega)$, with respect to these forces in accordance with the linear response theory:
\begin{equation}
\label{mu:def}
\delta s_{\alpha,\omega} = \sum_\beta \mu_{\alpha\beta}(\omega)
f_{\beta,\omega}\:.
\end{equation}
Thus, at high temperatures, $k_B T \gg \hbar\omega$, the spin noise spectrum can be recast as~\cite{ll5,glazov2018electron}
\begin{equation}
\label{fdt}
(\delta s_\alpha \delta s_\beta)_\omega = \frac{\mathrm i k_B T}{\omega}
\left[\mu_{\beta\alpha}^*(\omega) - \mu_{\alpha\beta}(\omega) \right]\:.
\end{equation}
This establishes relation between the spin noise spectroscopy and the electron spin resonance technique. We stress that this relation holds true in thermal equilibrium only. In non-equilibrium systems the susceptibility to external fields and noise spectra are, generally speaking, independent~\cite{springerlink:10.1007,glazov_keldysh}.

Also, the spin noise spectra can be expressed via the eigenfunctions and eigenenergies of the spin system~\cite{ll5}, see Sec.~\ref{sec:0D} and Eq.~\eqref{eq:spectrum_num} therein.

\subsection{Onsager relations}

Now let us analyze several general properties of the correlation functions. By definition, Eq.~\eqref{correlations:qnt}, the second order correlator $\mathcal C_{\alpha\beta}(\tau)$ satisfies the permutation relation
\begin{subequations}
\label{Onsager}
\begin{equation}
\label{ons:1}
\mathcal C_{\alpha\beta}(\tau) = \mathcal C_{\beta\alpha}(-\tau).
\end{equation}
In equilibrium, the time reversal symmetry leads to additional relations on the correlation functions~\cite{ll5}. In the absence of magnetic fields, it is of no importance which of the spin components  $\delta S_\alpha$ or $\delta S_\beta$ is taken at the earlier or later time moment in Eqs.~\eqref{correlations} and~\eqref{correlations:qnt}. Hence, at $\bm \Omega=0$ we have $\mathcal C_{\alpha\beta}(\tau) = \mathcal C_{\beta\alpha}(\tau) = \mathcal C_{\alpha\beta}(-\tau) = \mathcal C_{\beta\alpha}(-\tau)$. In the presence of magnetic field in addition to Eq.~\eqref{ons:1} we obtain
\begin{equation}
\label{ons:2}
\mathcal C_{\alpha\beta} (\tau; \bm \Omega) = \mathcal C_{\beta\alpha} (\tau; -\bm \Omega),
\end{equation}
\end{subequations}
because the magnetic field changes its sign at a time reversal $t\to-t$: $\bm \Omega \to -\bm \Omega$. Naturally, correlators in the frequency domain have the same properties. Moreover, since the replacement  $\omega\to-\omega$ corresponds to a complex conjugation, it follows then that
\begin{subequations}
\label{Onsager:frq}
\begin{equation}
\label{ons:1:frq}
\tilde{\mathcal C}_{\alpha\beta}(\omega) = \tilde{\mathcal C}_{\beta\alpha}^*(\omega) = \tilde{\mathcal C}_{\beta\alpha}(-\omega),
\end{equation}
\begin{equation}
\label{ons:2:frq}
\tilde{\mathcal C}_{\alpha\beta} (\omega; \bm \Omega) = \tilde{\mathcal C}_{\beta\alpha} (\omega; -\bm \Omega).
\end{equation}
\end{subequations}
Equations~\eqref{ons:2} and~\eqref{ons:2:frq} are known as Onsager relations. They describe the principle of symmetry of kinetic coefficients in thermal equilibrium conditions~\cite{PhysRev.37.405,ll5}. Evidently, Eqs.~\eqref{noise:field} satisfy the principle of symmetry of kinetic coefficients~\eqref{Onsager:frq}.

\subsection{Relation between the spin noise spectroscopy and Raman scattering of light}
\label{sec:Raman}

In optical experiments on spin noise spectroscopy, as discussed above, the spin noise is detected via fluctuations of the Faraday, Kerr and ellipticity effects. Fluctuations of the sample magnetization result in a modulation of the polarization of the monochromatic wave transmitted through the sample. As a result, the spectrum of electromagnetic field is enriched. This results in a relation between the fluctuations of the Faraday and Kerr rotation, as well as ellipticity, and the spin-flip Raman scattering of light~\cite{gorb_perel}.

\begin{figure*}[t]
\includegraphics[width=\linewidth]{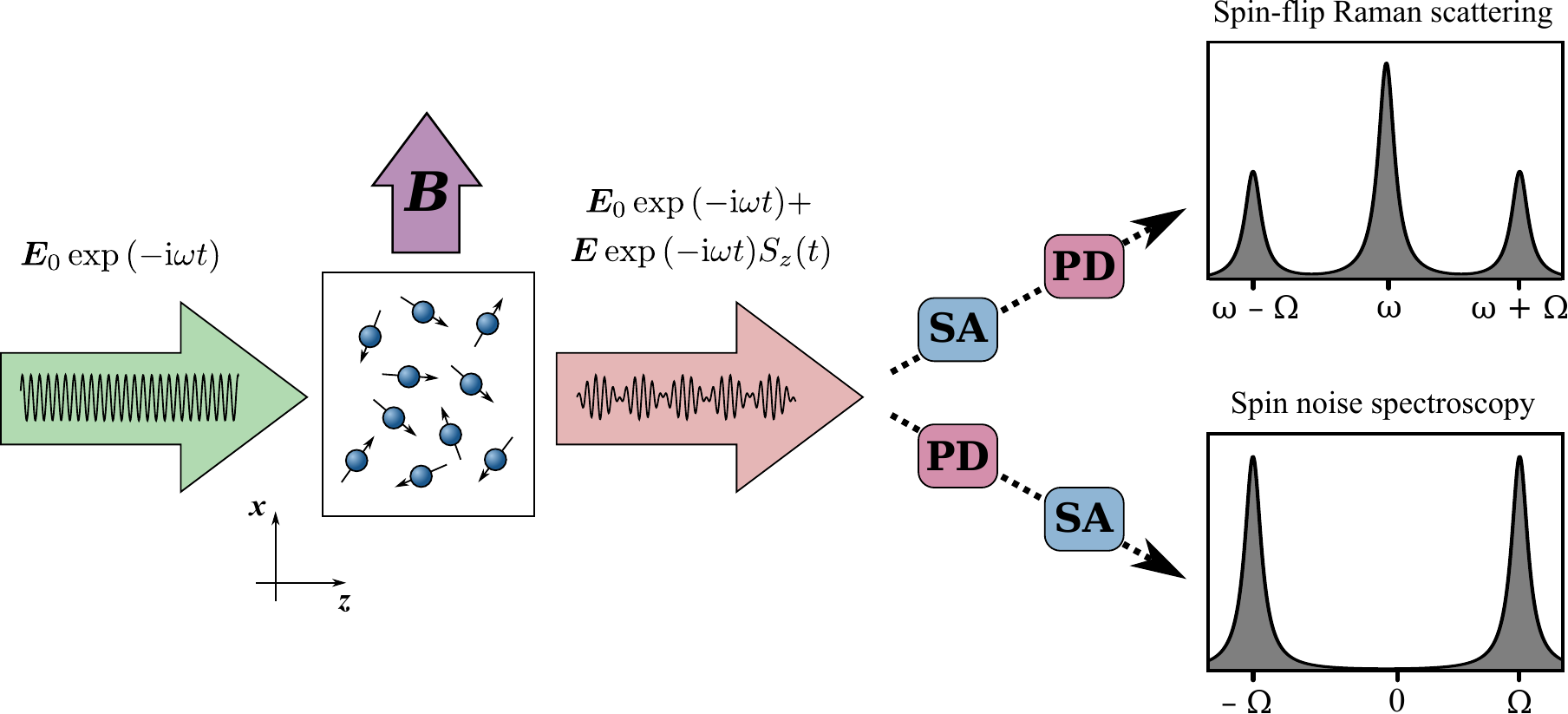}
\caption{
Transmission of the probe beam through the sample. ``SA'' denotes the spectrum analyzer and ``PD'' denotes the photodetector. In the spin-flip Raman scattering experiments the spectrum analyzer is used in the optical frequencies domain. In the spin noise spectroscopy it is used in the radio frequencies domain. (Adapted from Ref.~\cite{Glazov:15}.)
 }\label{fig:sec2:scatt}
\end{figure*}

Let us address these effects in more detail following Ref.~\cite{Glazov:15}. Let the probe beam polarized along the $x$-axis with the frequency $\omega$ propagate through a medium along the  $z$-axis. Then the contribution to the dielectric polarization of the medium caused by the spin fluctuation  $\delta S_z(t)$ can be written in the form~\cite{Glazov:15,glazov2018electron}
\begin{equation}
\label{dPy}
\delta P_y \propto \frac{\delta S_z(t) E_{0,x}\e^{-\i\omega t}}{\omega_0 - \omega - \mathrm i \gamma},
\end{equation}
where $E_{0,x}$ is the amplitude of the probe beam, $\omega_0$ and $\gamma$ are the eigenfrequency and damping of the resonance of the medium used to detect the spin fluctuations. It is assumed that the detuning $|\omega_0 - \omega|$ is small as compared to the distances to other resonances. We also assume that the spin fluctuations  $\delta S_z(t)$ are ``slow'' as compared to $\gamma^{-1}$. The presence of orthogonal to the initial polarization component $\delta P_y$ of the dipole moment results in the polarization plane rotation, with an instantaneous fluctuation of the Faraday rotation angle being [cf. Eq.~\eqref{dtheta:gen}]
\begin{equation}
\label{dtheta:res}
\delta \vartheta(t) \propto \Re\{ \delta P_y E_{0,x}^*\} \propto \delta S_z(t)  \frac{\omega - \omega_0}{(\omega-\omega_0)^2+\gamma^2}.
\end{equation}
In spin noise spectroscopy experiments, the correlator $\langle\delta\vartheta(t)\delta\vartheta(t')\rangle$ (or its Fourier-transform) is measured. To that end, the field transmitted through the sample is detected using a photodetector, which enables one to determine the light intensity and Stokes parameters, see Fig.~\ref{fig:sec2:scatt}. Next, the signal arrives at the spectrum analyzer where the Faraday rotation angle  spectrum (generally speaking, the spectrum of light intensity) is acquired~\cite{Glazov:15,:/content/aip/journal/rsi/87/9/10.1063/1.4962863}.

It follows from Eq.~\eqref{dPy} that the spectrum of the probe field is enriched due to the fluctuations  $\delta S_z(t)$. For example, in the simplest case, when a transverse magnetic field is applied,  $\delta S_z(t) \propto \cos(\Omega t)$ [cf. Eq~\eqref{eq:transverse}], spin fluctuations result in the appearance of the secondary or scattered waves with the frequencies  $\omega \pm \Omega$. This secondary field is usually detected in the inelastic spin-flip light scattering~\cite{gorb_perel,Glazov:15,:/content/aip/journal/rsi/87/9/10.1063/1.4962863}, where the field transmitted through the sample is first passed through a spectrometer and then converted to a photocurrent of a detector. As a result, the intensities of corresponding spectral components are detected, Fig.~\ref{fig:sec2:scatt}. The analysis above underlines a deep connection between the spin noise spectroscopy technique and spin-flip Raman scattering: Both methods allow one to determine the correlation function of spin fluctuations.


\section{Zero dimensional systems}
\label{sec:0D}

Semiconductor systems with localized charge carriers are among the most studied ones by the spin noise spectroscopy technique. The key and attractive feature of such systems is a long, up to micro- or even milliseconds, spin relaxation time in typical samples with quantum dots, that results from quenching of spin relaxation mechanisms  related to orbital motion of electrons. The main cause of the spin decoherence of localized change carriers is, as a rule, a hyperfine interaction with host lattice nuclear spins~\cite{glazov2018electron}. In this section we describe the spin noise spectra in systems with pronounced hyperfine interaction in moderate magnetic fields.

\subsection{Central spin model}
\label{sec:central}

The spin Hamiltonian of a localized electron in an external magnetic field $\bm B$ in the presence of the hyperfine interaction with the host lattice nuclear spins has the form
\begin{equation}
  \label{eq:simplest_ham}
  \mathcal H = \hbar\bm\Omega_L\bm s + \sum_{k=1}^{N_n}A_k \bm I_k\bm s.
\end{equation}
Here $\bm s$ is the electron spin operator,  $\bm\Omega_L=g_e\mu_B\bm B/\hbar$ is the Larmor precession frequency in the external magnetic field, $g_e$ is the effective electron $g$-factor (its anisotropy is neglected), subscript $k$ enumerates $N_n$ nuclear spins which effectively interact with the electron spin, $\bm I_k$ are the nuclear spin operators and  $A_k$ are the hyperfine coupling constants. The latter are determined both by the parameters of isotopes of the lattice and by the electron wavefunction~\cite{abragam_rus,glazov2018electron}. The magnetic field, as before, is assumed to be sufficiently weak and the temperature is assumed to be sufficiently large to neglect the average electron spin polarization. In thermal equilibrium the nuclear spins are oriented randomly, therefore the electron experiences a random nuclear field
\begin{equation}
  \label{eq:Omega_N}
  \hbar\bm\Omega_N=\sum_{k=1}^{N_n}A_k\bm I_k.
\end{equation}
In real III-V semiconductor systems the number of nuclei in a quantum dot is $N_n\sim 10^4$---$10^6$. The mean square fluctuation of the nuclear field is given by ${\left\langle\bm\Omega_N^2\right\rangle=3\delta_e^2/2}$, where the parameter $\delta_e$ is defined as
\begin{equation}
  \label{eq:delta_e}
  \delta_e^2 = \frac{2}{3}\sum_{k=1}^{N_n}I_k(I_k+1)A_k^2/\hbar^2.
\end{equation}
The description of the intertwined spin dynamics of electron and nuclei in the framework of Hamiltonian~\eqref{eq:simplest_ham} is known in the literature as the central spin model~\cite{Gaudin,Yang_2016}.

In the framework of the model described by Eq.~\eqref{eq:simplest_ham} the nuclear spin dynamics is driven by the electron spin. Generally, the nuclear spin dynamics in quantum dots can also be related to the quadrupole interaction, i.e, the splitting of nuclear spin levels due to strain, and nuclear spin precession in the external magnetic field. Corresponding characteristic times exceed by far the electron spin precession period in the field of nuclear fluctuation $\sim1/\delta_e$. This is because of the small nuclear $g$-factor and quadrupole splittings as well as the small ratio of the Knight and Overhauser fields of the order of $1/\sqrt{N_n}$. Due to the separation of the timescales, the nuclear spins at the times relevant for the electron spin dynamics can be considered as frozen~\cite{merkulov02}. In this case one can apply a ``semiclassical'' approach to describe the spin noise~\cite{NoiseGlazov}, which is based on the averaging of the dynamics of the electron spin fluctuations over the distribution of  $\bm\Omega_N$.

Hence, for a given direction of the nuclear field, the electron spin correlation functions are given by Eq.~\eqref{noise:field}, where $\bm\Omega=\bm\Omega_L+\bm\Omega_N$ is the total electron spin precession frequency. Due to the large number of nuclei, $N_n\gg 1$, and the statistical independence of their spins, the distribution function of nuclear fields is Gaussian~\cite{merkulov02}
\begin{equation}
  \label{eq:F_e}
  \mathcal F(\bm\Omega_N) = \frac{1}{(\sqrt{\pi}\delta_e)^3}\exp\left(-\frac{\Omega_{N}^2}{\delta_e^2}\right).
\end{equation}
In the case of holes the hyperfine interaction is anisotropic, so the dispersions of the Overhauser field as well as the effective  $g$-factors in different directions can differ~\cite{ivchenko05a,Urbaszek}. Making use of Eq.~\eqref{noise:field} one can readily calculate the spin noise spectrum per electron~\cite{NoiseGlazov}
\begin{equation}
  \label{eq:one_dot_noise}
  (\delta S_z^2)_\omega = \frac{\pi}{2} \int\limits\d\bm\Omega_N\mathcal F(\bm\Omega_N)
\left\lbrace\cos^2(\theta)\Delta(\omega)+\sin^2(\theta)\frac{\Delta(\omega-\Omega)+\Delta(\omega+\Omega)}{2}\right\rbrace,
\end{equation}
where $\theta$ is an angle between $\bm\Omega$ and the $z$ axis and $\tau_s$ in  $\Delta(x)$ [see Eq.~\eqref{eq:Delta_def}] is a phenomenological spin relaxation time unrelated to the hyperfine interaction. In the ``box'' model where all hyperfine coupling constants are equal, the number of nuclear spins is large and an external field is absent, Eq.~\eqref{eq:one_dot_noise} is exact~\cite{Bortz_2007,Kozlov2007}.

\begin{figure}[htpb]
\center{
\includegraphics[width=0.7\textwidth]{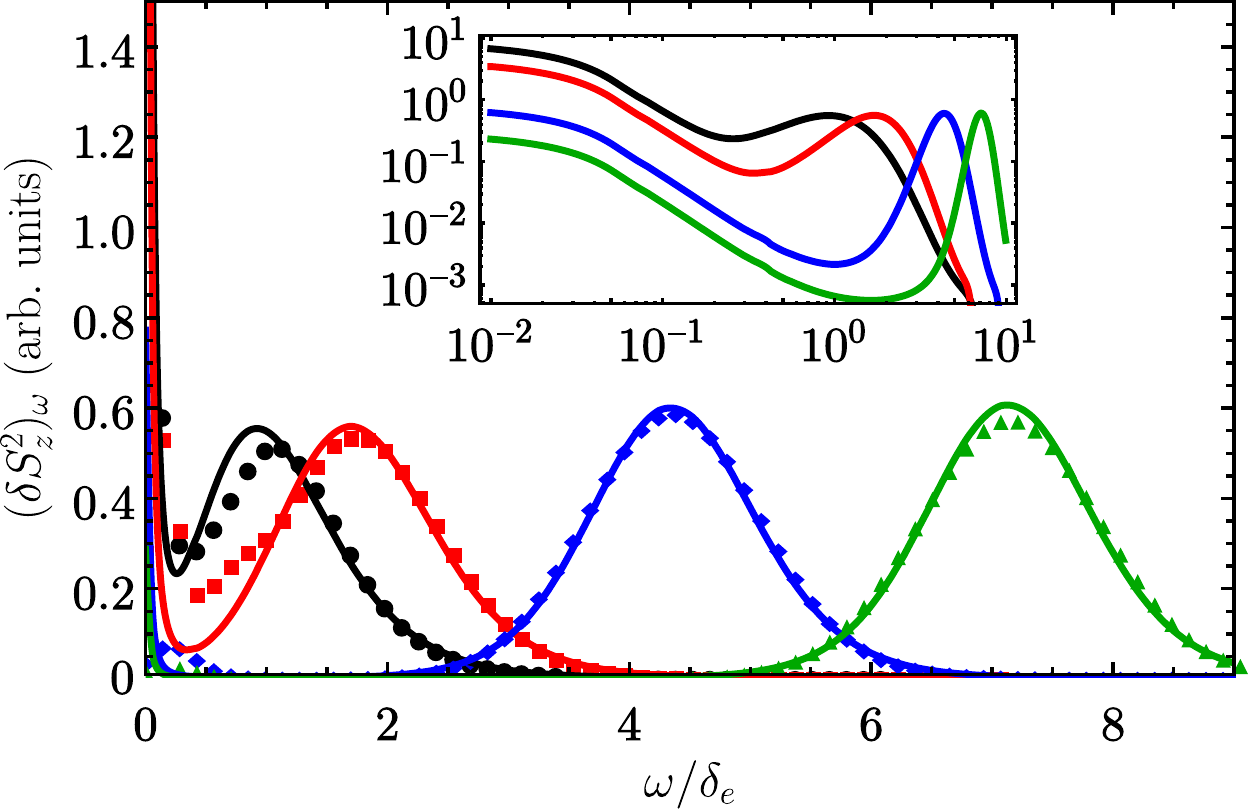}
\caption{\label{fig:num_anal} Spin noise spectra calculated after Eq.~\eqref{eq:one_dot_noise} (solid curves) and numerically in the central spin model accounting for the nuclear spin dynamics (dots) for different transverse external fields  $\Omega_L=0$ (black), $\Omega_L=\sqrt{2}\delta_e$ (red), $\Omega_L=3\sqrt{2}\delta_e$ (blue), $\Omega_L=5\sqrt{2}\delta_e$ (green) for $\tau_s\delta_e=25\sqrt{2}$. Inset shows the solid curves in the bilogarithmic scale.
}
}
\end{figure}

As a rule, the spin relaxation time is much longer than the spin precession period in the nuclear field, i.e, the condition $\delta_e\tau_s\gg 1$ holds. In this case it is possible to derive an analytical expression for the spin noise spectrum in zero magnetic field:
\begin{equation}
  \label{eq:two_peaks}
  (\delta S_z^2)_\omega = \frac{\pi}{6}\Delta(\omega)+\frac{2\sqrt{\pi}\omega^2}{3\delta_e^3}\e^{-\omega^2/\delta_e^2},
\end{equation}
which is shown by the black curve in Fig.~\ref{fig:num_anal}. The spectrum consists of two peaks. The first peak at the zero frequency is very high and narrow. It corresponds to the relaxation with the time constant $\tau_s$ of the spin component parallel to $\bm\Omega_N$. This peak is denoted as relaxational. The second peak with the maximum at  $\omega=\delta_e$ corresponds to the electron spin precession in the random nuclear field, its shape reflects the distribution function of the absolute value $\Omega_N$ and its width is of the order of $\delta_e$.

Note that the spectral weight of the zero frequency peak is  $1/3$ of the total intensity of the spin noise. This is a result of the isotropic distribution of hyperfine fields. Application of the transverse external magnetic field reduces the probability to find the total field $\bm\Omega$ parallel to the $z$ axis. This probability is proportional to  $\cos^2\theta$. As a result, the peak at the zero frequency gets suppressed while the precession peak shifts to the higher frequencies, as shown in Fig.~\ref{fig:num_anal}. In the high-field limit, $\Omega_L\gg\delta_e$, the spin noise spectrum has the form
\begin{equation}
  \label{eq:gauss_peak}
  (\delta S_z^2)_\omega = \frac{\sqrt{\pi}}{4\delta_e}\e^{-(\omega-\Omega_L)^2/\delta_e^2}.
\end{equation}
It follows from Eq.~\eqref{eq:gauss_peak} that the spectrum is Gaussian centered at  $\Omega_L$ with the width determined by a typical fluctuation of the Overhauser field. For electrons localized in quantum dots, as a rule, the spread of $g$ factors is present. It results in the additional broadening of the spectrum which linearly increases with increase of the field~\cite{crooker2010}.

The effect of the longitudinal magnetic field is opposite. It results in the decrease of the typical values of the angle  $\theta$ in Eq.~\eqref{eq:one_dot_noise}, i.e., in the suppression of the precessional peak and enhancement of the relaxational one. In the strong magnetic field limit, $\Omega_L\gg\delta_e$, the spin noise spectrum is Lorentzian
\begin{equation}
  \label{eq:one_peak}
  (\delta S_z^2)_\omega = \frac{\pi}{2}\Delta(\omega),
\end{equation}
and it is centered at $\omega=0$. Its height is three times larger as compared with the case of zero field. In quantum dot ensembles, the shape of the peak differs from Lorentzian, which corresponds to non-monoexponential spin relaxation in the longitudinal magnetic field~\cite{PhysRevLett.108.186603}. In the single quantum dot, the zero-frequency peak is well described by the Lorentzian function~\cite{singleHole}.

The model described above was successfully applied for description of the spin noise and determination of the spin parameters in quantum dot ensembles doped with electrons and holes~\cite{eh_noise}.

The results presented above were derived in the model where the nuclear spin dynamics is neglected. The central spin model can account for the nuclear spin precession in the Knight field. Generally, the spin noise spectrum can be calculated as
\begin{equation}
  \label{eq:spectrum_num}
  (\delta S_z^2)_\omega = \frac{2\pi}{\mathcal N}\sum_{n,m}\left\langle n\left| S_z \right|m\right\rangle\left\langle m\left| S_z \right|n\right\rangle\Delta\left(\omega-\frac{E_n-E_m}{\hbar}\right),
\end{equation}
where $n,m=1\ldots\mathcal N$ enumerate eigenstates of the Hamiltonian~\eqref{eq:simplest_ham}, $E_n$ and $E_m$ are their energies.

Despite the fact that the central spin model can be exactly solved by Bethe ansatz~\cite{Gaudin}, the actual calculations are most efficiently performed in the time representation by decomposition of the evolution operator  $\exp(-\i\mathcal H \tau/\hbar)$ over the Chebyshev polynomials~\cite{anders-CPT}. It ensures a uniform convergence regardless the initial state of the system. Since the number of the eigenstates  $\mathcal N$ exponentially increases with increase of the number of nuclei $N_n$, in actual calculations it is possible to address only several tens of nuclei (apart from the case of the simplified ``box'' model~\cite{Bortz_2007,Kozlov2007}). An advantage of the numerical approach is the possibility to exactly account for the quadrupole splitting of nuclear spin sublevels~\cite{hackmann2015}.

The results of numerical calculations are shown by dots in Fig.~\ref{fig:num_anal}. It is seen that the shape of precessional peak almost coincides in both approaches. The agreement becomes better with increase of the number of nuclear spins in the numerical modelling. The agreement is almost complete at the transverse magnetic field $\Omega_L\gtrsim\delta_e$.

From experimental point of view, the peaks broadening due to the hyperfine interaction and  $g$-factor distribution can be considered as a detrimental for applications effect. In Refs.~\cite{braun2007,PhysRevB.91.155301} an application of an alternating magnetic field to overcome it has been suggested. It results in the series of narrow peaks with the width $1/\tau_s$ at the frequencies determined by the field modulation frequency. If, in addition to the alternating field, the system is subject to a static magnetic field, the spin noise spectra demonstrate the Mollow triplet structure~\cite{PhysRevB.101.075403}. It can be used to measure the spin relaxation time $\tau_s$ in the conditions when it is long as compared to the spin dephasing time $T_2^*\sim1/\delta_e$. This kind of experiments have been performed for potassium atoms~\cite{Mollow-noise} and it is perspective to perform similar studies for quantum dots.

\begin{figure}
  \centering
  \includegraphics[width=0.6\linewidth]{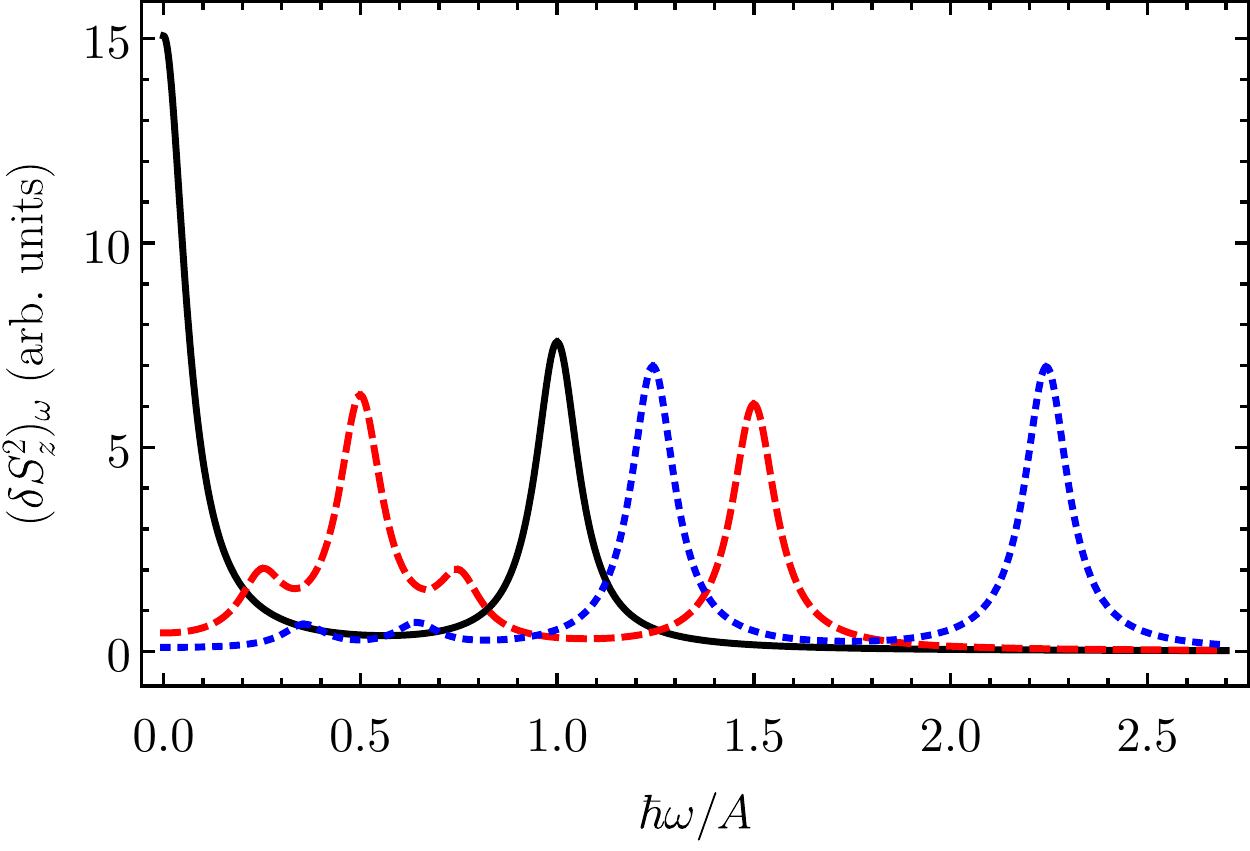}
  \caption{Spin noise spectrum of the electron interacting with a donor spin  $I=1/2$ in external magnetic fields $\hbar\Omega_L/A=0$ (solid black curve), $0.75$ (dotted red curve), and $1.6$ (dotted blue curve), calculated from the correlation function Eq.~\eqref{eq:corr1/2}.}
  \label{fig:1/2}
\end{figure}

In addition to the problem of calculation of the spin noise of an electron interacting with a large number of nuclei, one can analytically describe spin fluctuations of an electron interacting with a single nuclear spin  $\bm I$. This situation is realized, for example, for donor-bound electrons in isotopically purified II-VI semiconductors~\cite{ZnOnoise,PhysRevB.100.205415}. Since the isotropic hyperfine interaction conserves the total spin, it is convenient to project the Hamiltonian of the system onto a subspace, where the total spin component along the magnetic field axis is $I_{z_1}+S_{z_1}=m+1/2$. For $m=-I\ldots I-1$ this ``shortened'' Hamiltonian can be written as (for the states with $m=-I-1$ and $I$ spin dynamics is absent)
\begin{equation}
  \mathcal H_m=\hbar(\bm\Omega_m+\bm\Omega_L)\bm S-A/4,
\end{equation}
where $\hbar\Omega_m\equiv\hbar\Omega=(I+1/2)A$ is the splitting between the states with the total momenta $I\pm1/2$, $A$ is the hyperfine coupling constant, and the direction of the vector $\bm\Omega_m$ is determined by its component $\hbar\Omega_{m,z}=(m+1/2)A$. Here, as above, we neglect a Zeeman splitting of nuclear spin sublevels. For zero magnetic field, after averaging over all values of $m$, one finds the spin correlation function in the following form
\begin{equation}
  \braket{S_z(0)S_z(t)}=\frac{\e^{-t/\tau_s}}{12(2I+1)^2}\left[4I^2+4I+3+8I(I+1)\cos(\Omega t)\right].
\end{equation}
The corresponding spin noise spectrum consists of two peaks at  $\omega=0$ and $\omega=\Omega$. For a transverse magnetic field for $I=1/2$ the spin correlator has the form
\begin{equation}
  \label{eq:corr1/2}
  \braket{S_z(0)S_z(t)}=\frac{\e^{-t/\tau_s}}{4}\cos\left(\frac{\Omega t}{2}\right)\left[\cos\left(\frac{\Omega_L t}{2}\right)\cos\left(\frac{\tilde\Omega t}{2}\right)-\frac{\Omega_L}{\tilde\Omega}\sin\left(\frac{\Omega_L t}{2}\right)\sin\left(\frac{\tilde\Omega t}{2}\right)\right],
\end{equation}
where $\tilde\Omega=\sqrt{\Omega_L^2+(A/\hbar)^2}$. The corresponding spin noise spectrum is shown in Fig.~\ref{fig:1/2}. Generally, it consists of four peaks at the frequencies $\omega=|\Omega\pm\Omega_L\pm\tilde\Omega|/2$. In the strong field limit  $\Omega_L\gg\Omega$, the spectrum consists of two peaks at  $\Omega_L\pm\Omega/2$~\cite{PhysRevB.97.195311}. Weak magnetic field shifts the relaxational peak towards the frequency $\Omega_L/2$, which corresponds to an effective $g$-factor $g_e/2$. Indeed, a weak field does not mix the singlet and triplet states of the electron-nucleus system, thus the total spin precesses with twice smaller frequency than that of the electron. In the general case of $I>1/2$ the small-field effective $g$-factor equals to $g_e/(2I+1)$~\cite{ll3}.

\subsection{Effect of exchange interaction}
\label{sec:exch}

Manybody effects can be prominent in ensembles of localized electrons. Let us consider, for example, an ensemble of donor-bound electrons in bulk GaAs-type semiconductor. A typical magnitude of the electron spin precession frequency is $\delta_e \sim 2 \times 10^8$~s$^{-1}$~\cite{Dzhioev:2002kx,Romer2010}. Electron-electron exchange interaction becomes comparable with the hyperfine interaction at the distances between the donors of about  $0.1~\mu$m, which corresponds to the donor density   $n_d=10^{14} \div 10^{15}$~cm$^{-3}$. Since in GaAs the metal-insulator transition takes place at much higher donor densities $\sim2\times 10^{16}$~cm$^{-3}$~\cite{Dzhioev02}, one has to take into account an interplay between the electron-electron exchange interaction and hyperfine interaction between electron and nuclear spins to describe the spin noise of donor-bound electrons in bulk semiconductor even at relatively low doping level. Theoretically, this interplay from the point of view of spin noise has been studied in Ref.~[\onlinecite{noise-exchange-rus}].

\begin{figure}
  \centering
  {\includegraphics[width=0.5\linewidth]{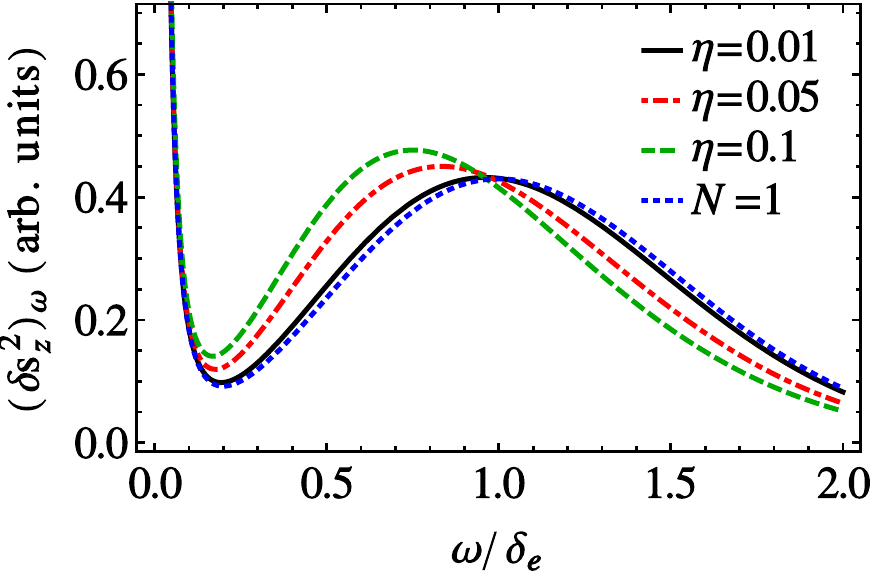}}
  \caption{\label{fig:exchange}Spin noise spectrum of the donor ensembles with different dimensionless densities $\eta=\pi n_d R_c^3/6$. Dotted blue curve was calculated neglecting the exchange interaction. (Adapted from Ref.~\cite{noise-exchange-rus}.)}
\end{figure}

A model of clusters has been proposed, where the ensemble of donors is divided into groups where the exchange interaction is more efficient then the hyperfine one. Inside the cluster, spin dynamics is controlled both by the exchange and hyperfine interaction, while the exchange interaction between the clusters can be neglected. One can say, that a manybody localization of spin excitations at clusters takes place.

A competition of the hyperfine and exchange interaction effects can be most simply described for the exampling case of two electrons strongly bound by the exchange interaction. Neglecting the hyperfine interaction, the states of the pair of electrons are characterized by the total spin  $S=0$ (singlet) and $S=1$ (triplet) and spin component  $S_z$ along a given axis $z$. If the splitting between the singlet and triplet is large enough, the mixing of these states via the hyperfine interaction can be neglected. One can, therefore, consider separately the dynamics of the triplet state with the total spin $S=1$. Fluctuations $\delta \mathbf S$ of the triplet spin state are described by Eq.~\eqref{ch9:field} with the effective nuclear field
\begin{equation}
\label{omega:triplet}
{\mathbf \Omega}_{\rm eff}=\frac{\mathbf \Omega_1 + \mathbf \Omega_2}{2}.
\end{equation}
The dispersion ${\mathbf \Omega}_{\rm eff}$ is two times smaller than that of the nuclear fields $\bm \Omega_1$ and $\bm \Omega_2$, acting on two electrons. Hence, the spin noise spectrum normalized per electron is given by Eq.~\eqref{eq:two_peaks} with the replacement  $\delta_e$ by $\delta_e/\sqrt{2}$. As a result, the peak at $\omega=0$ is the same as in the absence of the exchange interaction, while the peak resulting from the spin precession is shifted to the frequency $\delta_e/\sqrt{2}$, its height is $\sqrt{2}$ times increased and width, respectively, $\sqrt{2}$ times decreased as compared with those of the peak in the absence of the exchange interaction. This is due to an effective averaging of the nuclear fields caused by the exchange interaction.

In the ensemble of donors, an increase of the electron density characterized by a dimensionless parameter $\eta=\pi n_d R_c^3/6$, where $R_c$ is the distance between the donors with the exchange interaction equal to $\hbar\delta_e$, the precessional peak in the spin noise spectrum shifts to the lower frequencies, see Fig.~\ref{fig:exchange}. It is in qualitative agreement with the experimentally observed increase of the spin relaxation time with increase of the density of the donor-bound electrons in bulk GaAs~\cite{Dzhioev02}.

At sufficiently high density of the donors, the cluster model described above can not be applied. In this case, an infinite cluster is formed in the system, which corresponds, in the classical description, to a percolation, and in quantum description to a  delocalization of spin excitations by manybody interactions. The spin diffusion takes place over the infinite cluster, which due to hyperfine fields results in the spin relaxation ~\cite{KKavokin-review}. The spin noise spectrum in this situation consists of a single zero-frequency peak with the width controlled by the spin relaxation time. Modification of the spin noise spectra with increase of the donor density is discussed in more detail below in Sec.~\ref{sec:tun}.

\subsection{Effect of electron hopping}
\label{sec:hop}

\begin{figure}
  \centering
  \includegraphics[width=0.4\linewidth]{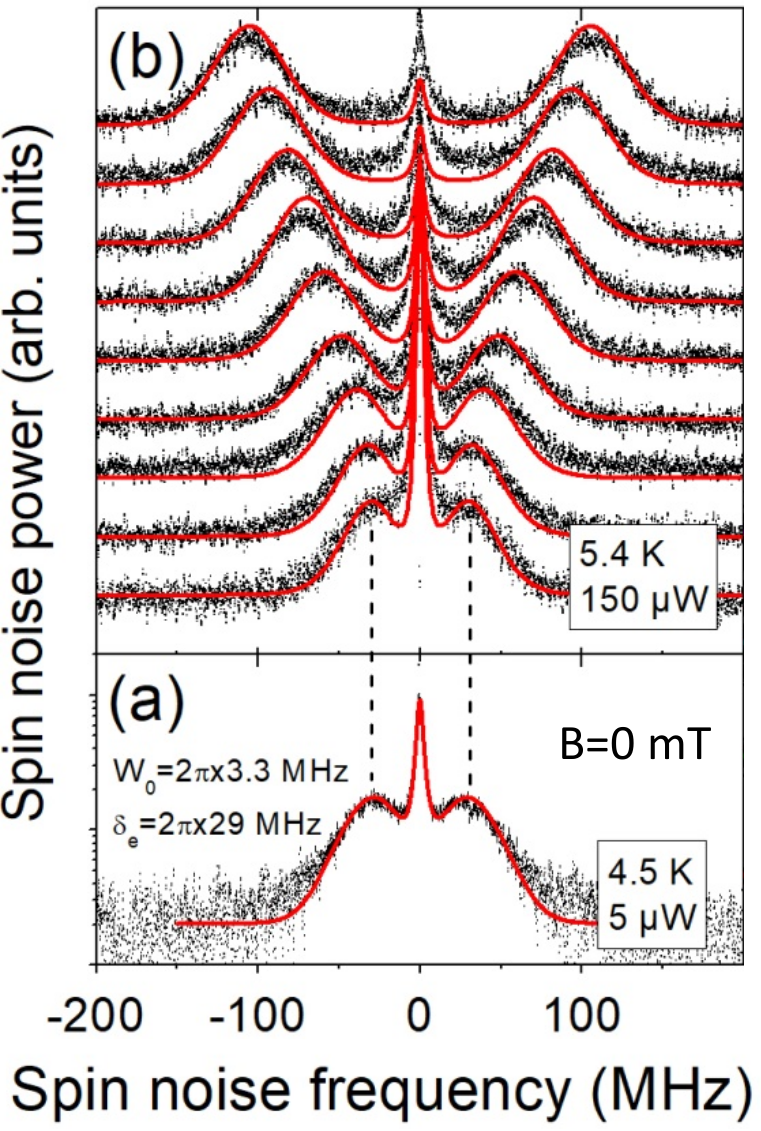}
  \caption{Spin noise spectra in  $n$-doped CdTe (a)~at a zero magnetic field and  (b)~at a field from $0$ to $4$~mT with the steps of $0.5$~mT (spectra are offset vertically for clarity). Red curves demonstrate the fit using the model of Eq.~\eqref{eq:hopping_kin}. (Adapted from Ref.~\cite{cronenberger2019spatiotemporal}.)}
  \label{fig:hopping}
\end{figure}

Apart from the exchange interaction, the electron hopping between the localization sites can also result in quantitative and even qualitative modification of the spin noise spectra. When only a small fraction of localization centers are occupied with electrons, the spin noise spectrum can be found from a single-particle kinetic equations:
\begin{equation}
  \label{eq:hopping_kin}
  \dot{\bm S}_i=\bm\Omega_i\times\bm S_i+\sum_j\left(W_{ij}\bm S_j-W_{ji}\bm S_i\right)-\bm S_i/\tau_s,
\end{equation}
where $\bm S_i$ is the average spin at the $i$th site, $\bm\Omega_i$ is the corresponding precession frequency in a field of the nuclei, and $W_{ij}$ are the hopping rates between the centers. According to a general rule, the correlation functions $\braket{\delta S_{i,\alpha}(t+\tau)\delta S_{j,\beta}(t)}$ satisfy the same set of equations~\cite{ll10}. It should be solved together with initial conditions describing the same-time correlators, see Sec.~\ref{sec:methods}. This set of equations can be solved analytically provided that the rates of all transitions are equal to each other: $W_{ij}=W_0/N$, where $W_0$ is the rate of the electron departure from a given site, $N$ is the total number of sites. The spectrum of the total spin, $\bm S=\sum_i\bm S_i$, fluctuations per electron in this model has the form~\cite{Glazov_hopping}
\begin{equation}
  (\delta S_z^2)_\omega=\frac{\tau_\omega}{4}\frac{\mathcal A(\tau_\omega)}{1-W_0\tau_\omega\mathcal A(\tau_\omega)}+\mbox{c.c.},
\end{equation}
where $1/\tau_\omega=1/\tau_s+W_0-\i\omega$ and
\begin{equation}
  \mathcal A(\tau)=\braket{\frac{1+\Omega_{i,z}^2\tau^2}{1+\Omega_i^2\tau^2}},
\end{equation}
with the averaging performed over all sites. This expression is valid both in the absence of an external field and when the field is applied along the  $z$ axis. Expressions for the spin noise in the transverse field are more cumbersome, they are given in Ref.~\cite{Glazov_hopping}.

When the field is absent, the averaging with the distribution function~\eqref{eq:F_e} can be done analytically with the result:
\begin{equation}
\label{A:tau:mg}
  \mathcal A(\tau)=\frac{1}{3}+\frac{4}{3(\delta_e\tau)^2}-\frac{4\sqrt{\pi}\e^{1/(\delta_e\tau)^2}}{3(\delta_e\tau)^3}\erfc\left(\frac{1}{\delta_e\tau}\right).
\end{equation}
Depending on the ratio $W_0/\delta_e$ the shape of the spectrum can be qualitatively different as will be discussed in more details in Sec.~\ref{sec:tun}. This model well describes the Si donor-bound electron spin noise in GaAs~\cite{NuclearNoise} and the Al donor-bound electron spin noise in CdTe~\cite{cronenberger2019spatiotemporal}, as illustrated in Fig.~\ref{fig:hopping}.

The model with $N\gg1$ described above neglects the effects of electron return to the initial localization center. These effects can be easily described for a pair of localized electrons~\cite{PhysRevB.100.075409}. The spin noise spectra for two electrons with account for the exchange and hyperfine interactions as well as the electron hopping between the sites have a universal low-frequency divergence $\sim\ln(1/\omega)$. It is caused by the competition of the spin blockade effect~\cite{Ono1313,PhysRevB.72.165308} and nuclei-induced electron spin precession. Indeed, if for two electrons the frequencies of spin precession are parallel, $\bm\Omega_1\parallel\bm\Omega_2$, then in the triplet state where the electron spins are parallel to the axis of frequencies, the electron hops are forbidden and the spin relaxation (at $\tau_s \to \infty$) is absent. If the angle between the hyperfine fields equals to $\theta$ and is small, the spin fluctuations decay due to the electron hopping between the sites as~\cite{Shumilin2015}
\begin{equation}
  \delta\bm S(t)\propto\e^{-\gamma[1-\cos(\theta)]t/2},
\end{equation}
where $\gamma$ is the hopping rate in the singlet state. Averaging of this expression over the directions of nuclear fields results in the long-time asymptotic behavior $\propto 1/t$, which corresponds to the logarithmic divergence of the spectra at low frequencies. For a small number $N$ of interacting electrons, the same arguments demonstrate that the spin fluctuations decay as  $1/t^{N-1}$, thus the noise amplitude at low frequencies remains finite even at $\tau_s \to \infty$. Hence, the effects of electron returns result, generally speaking, in qualitative modification of the spin noise spectra at low frequencies. Note that there is an exponentially wide spread of the transition rates $W_{ij}$ in real systems, which may result in additional features of low-frequency spin fluctuations~\cite{Shumilin2015}.

\subsection{Effect of quantum-mechanical tunneling}
\label{sec:tun}

The spin noise spectra can be modified by the electron tunneling between the localization centers apart from the effects of the exchange interaction and hopping. In this section we assume, for simplicity, that the electron density is much lower than the density of the localization sites. Thus, we neglect the electron-electron interaction. We also assume that inelastic hopping is absent, e.g., due to the low temperature. In this situation the system Hamiltonian has the form
\begin{equation}
  \mathcal H={\sum_{i,j,\sigma}t_{ij}c_{i,\sigma}^\dag c_{j,\sigma}}+\hbar\sum_i\bm\Omega_i\bm S_i,
\end{equation}
where $t_{ij}$ are the tunneling matrix elements between the centers,  $c_{i,\sigma}$ is the annihilation operator of the electron in the spin state $\sigma$ at the site $i$. The spin noise spectrum in this case can be calculated using Eq.~\eqref{eq:spectrum_num}.

\begin{figure}
  \centering
  \includegraphics[width=\linewidth]{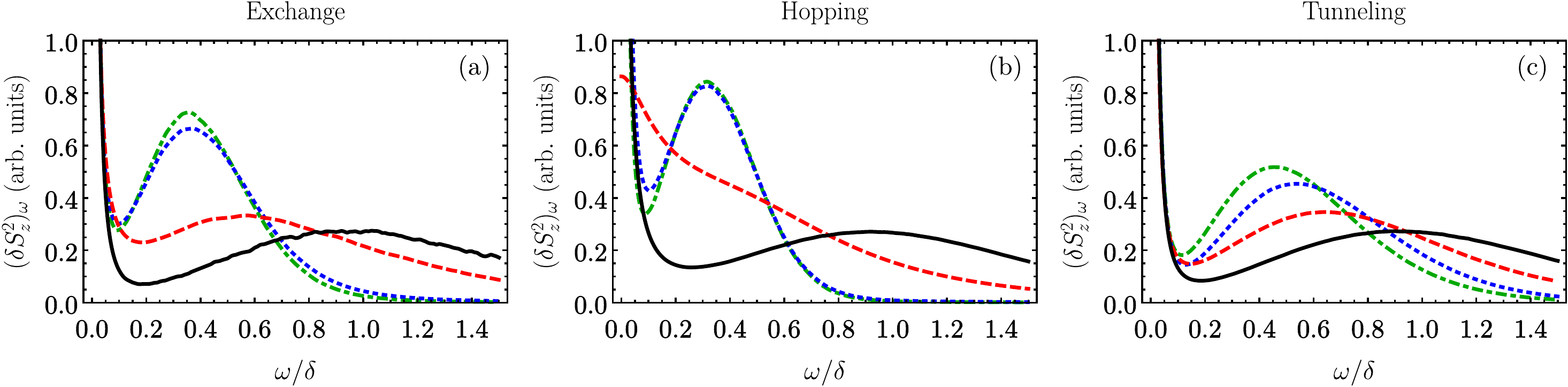}
  \caption{Spin noise spectra of localized electrons calculated accounting for (a) the exchange interaction (Sec.~\ref{sec:exch}), (b) hopping (Sec.~\ref{sec:hop}), and (c) quantum-mechanical tunneling (Sec.~\ref{sec:tun}). The parameters of calculation are $J_0/(\hbar\delta)=W_0/\delta=t_0/(\hbar\delta)=10^3$ and $\tau_s\delta=100$, the density of centers  $na_B^3=10^{-4}$ (black), $10^{-3}$ (red), $10^{-2}$ (blue), and $10^{-1}$ (green).}
  \label{fig:models}
\end{figure}

It is instructive to compare the modifications of the noise spectra in the models described above, which account for an exchange interaction, electron hopping and tunneling. To that end, we have calculated the spin noise spectra for an ensemble of $N=10$ randomly and independently distributed sites with different densities using the periodic boundary conditions. For the three models under consideration we have chosen the analogous set of the parameters: (a) exchange interaction constants $J_{ij}=J_0\e^{-r_{ij}/a_B}$, (b) hopping rates $W_{ij}=W_0\e^{-r_{ij}/a_B}$, and (c) tunneling constants $t_{ij}=t_0\e^{-r_{ij}/a_B}$, with $J_0/(\hbar\delta)=W_0/\delta=t_0/(\hbar\delta)\gg 1$. The results of the calculations are presented in Fig.~\ref{fig:models}.

Since the characteristic distance between the centers is $l=n^{-1/3}$, the shape of the spectra is determined by a dimensionless parameter   $\xi=J_0\e^{-l/a_B}/(\hbar\delta)=W_0\e^{-l/a_B}/\delta=t_0\e^{-l/a_B}/(\hbar\delta)$. For the black curves in Fig.~\ref{fig:models} this parameter is small ($\xi\sim10^{-17}$). In this case the hyperfine interaction of electron and nuclear spins is much stronger than the coupling with the other centers. The spin noise spectrum is, therefore, described by Eq.~\eqref{eq:two_peaks}. With increase of the density, the spin gets distributed over several centers. In the case of the exchange interaction [Fig.~\ref{fig:models}(a)], as described in Sec.~\ref{sec:exch}, a manybody localization at small clusters takes place~\cite{Parameswaran_2018,RevModPhys.91.021001}, and afterwards the spin becomes distributed over all $N$ sites. In this case, the nuclear fields are effectively averaged, so the spin precession peak shifts to lower frequencies. As noted above, the spin noise spectrum consists of a single peak at zero frequency in the limit of infinite number of localization sites and their high density.

In the case of hopping, the spin fluctuations are also get delocalized. Qualitatively, the electron spin precesses in the effective magnetic field, which changes with time as a result of electron hopping between the centers with different $\bm \Omega_{N,i}$. If the hops are inefficient,  $W_0/\delta_e \ll 1$, they only broaden the zero frequency peak. In this case Eq.~\eqref{eq:two_peaks} holds with the replacement $\tau_s^{-1} \to \tau_s^{-1} + 2W_0/3$. At a higher density of centers and higher temperature when the hopping becomes fast, $W_0/\delta_e \gg 1$, the dynamical averaging of nuclear fields takes place. In the infinite system, $N\to\infty$, the spin noise spectrum has a Lorentzian shape with the width $\sim \delta_e^2/W_0$. Note that for the green curve in Fig.~\ref{fig:models}(b) the electron travel time of all $N$ centers is shorter than the spin precession period, thus the fields of all the centers are averaged.

The case of quantum-mechanical tunneling is qualitatively different from those described above. This is because even if the parameter $\xi$ is not small, the electrons are localized by the Lifshits mechanism~\cite{Lifshits:1964_rus,Efros89}. The origin is the exponentially broad distribution of the tunneling constants (strong ``off-diagonal'' disorder). With increase of the density the localization length slowly increases and the spin precession peak shifts to lower frequencies. However, in contrast to the previous cases, the electron delocalization takes place at much higher densities, when  $na_B^3\sim1$. This regime is not reached for the parameters of calculations used for Fig.~\ref{fig:models}.

Interestingly, in the model where the nuclear fields are absent, but the exchange interaction between the localized electrons is present having the exponentially broad distribution of the coupling constants, the manybody localization of the spin fluctuations due to this spread (i.e., by the mechanism analogous to the Lifshits model) does not take place. The localization is prevented by the conservation of the total angular momentum. It is clear, for example, that the state with the maximal total spin of all electrons is manybody delocalized ~\cite{Shumilin_2016}. The analysis of the symmetries of the Hamiltonian shows that the system in this case is ergodic~\cite{Parameswaran_2018}. If the total angular momentum conservation law is broken, e.g., in the case of anisotropic exchange interaction, the system allows for a manybody localization. Such situations are analyzed in more detail in the next Sec.~\ref{sec:1D} in respect to the fluctuations in one-dimensional spin chains.

\subsection{Spin noise in non-equilibrium conditions}
\label{sec:4level}

Initially, the spin noise spectroscopy was considered as a tool to study the spin properties of a system in the conditions close to the thermal equilibrium~\cite{muller-Wells,PhysRevLett.108.186603}. Indeed, the fluctuation dissipation theorem is applicable in this case, and the spin noise spectrum reflects the frequency dependence of the imaginary part of the spin susceptibility, see Eq.~\eqref{fdt}. If the system is brought out of equilibrium, e.g., in the conditions of optical orientation of spins, the fluctuation dissipation theorem is not applicable. This situation requires development of the corresponding microscopic theory of non-equilibrium spin noise analogous to that of non-equilibrium fluctuations of electric current~\cite{Lax1,glazov_keldysh,springerlink:10.1007,Shulman_rus,Kogan}. In this case, the spin noise spectroscopy allows one to obtain more information about the systems dynamics as compared to the equilibrium conditions~\cite{Mollow-noise}. For bulk semiconductors under optical orientation conditions such problem has been solved in the pioneering work~\cite{ivchenko73fluct}.

As we have already demonstrated, the spin dynamics of localized charge carriers is mainly controlled by hyperfine interaction with the host lattice nuclei. Naturally, a question arises about the modification of the spin noise spectra in the case, when the nuclear spin system is out of thermodynamic equilibrium. The simplest example is the  non-equilibrium nuclear spin polarization. Experimentally, it can be achieved using the dynamic polarization of nuclei by circularly polarized light, which will be analyzed in more detail in Sec.~\ref{sec:extended}. Here we only note that a considerable (exceeding 50\%) nuclear spin polarization suppresses the nuclear spin fluctuations and modifies the distribution function of nuclear spins $\mathcal F(\bm \Omega_N)$~\eqref{eq:F_e}~\cite{Gangloff62}. As a result, the peak in the electron spin noise spectrum can narrow down as compared with the result of Eq.~\eqref{eq:gauss_peak}~\cite{polarizednuclei}. Its width can be reduced by several orders of magnitude, if the nuclear spin polarization approaches 100\%.

The electron spin precession mode-locking effect in the external magnetic field arising at the optical orientation by a train of circularly polarized pulses~\cite{A.Greilich09282007,yugova11} can serve as another example. The nuclear spins get tuned in such a way, that the electron spin precession becomes commensurable with the pulse repetition rate. As a result, the electron spin noise spectrum takes the shape of a sequence of narrow peaks at the frequencies satisfying this synchronization condition~\cite{PhysRevB.98.045307}. Such a structure of the spin noise spectrum directly reveals non-equilibrium nuclear fields distribution function.

\begin{figure}
\center{
  \includegraphics[width=0.6\linewidth]{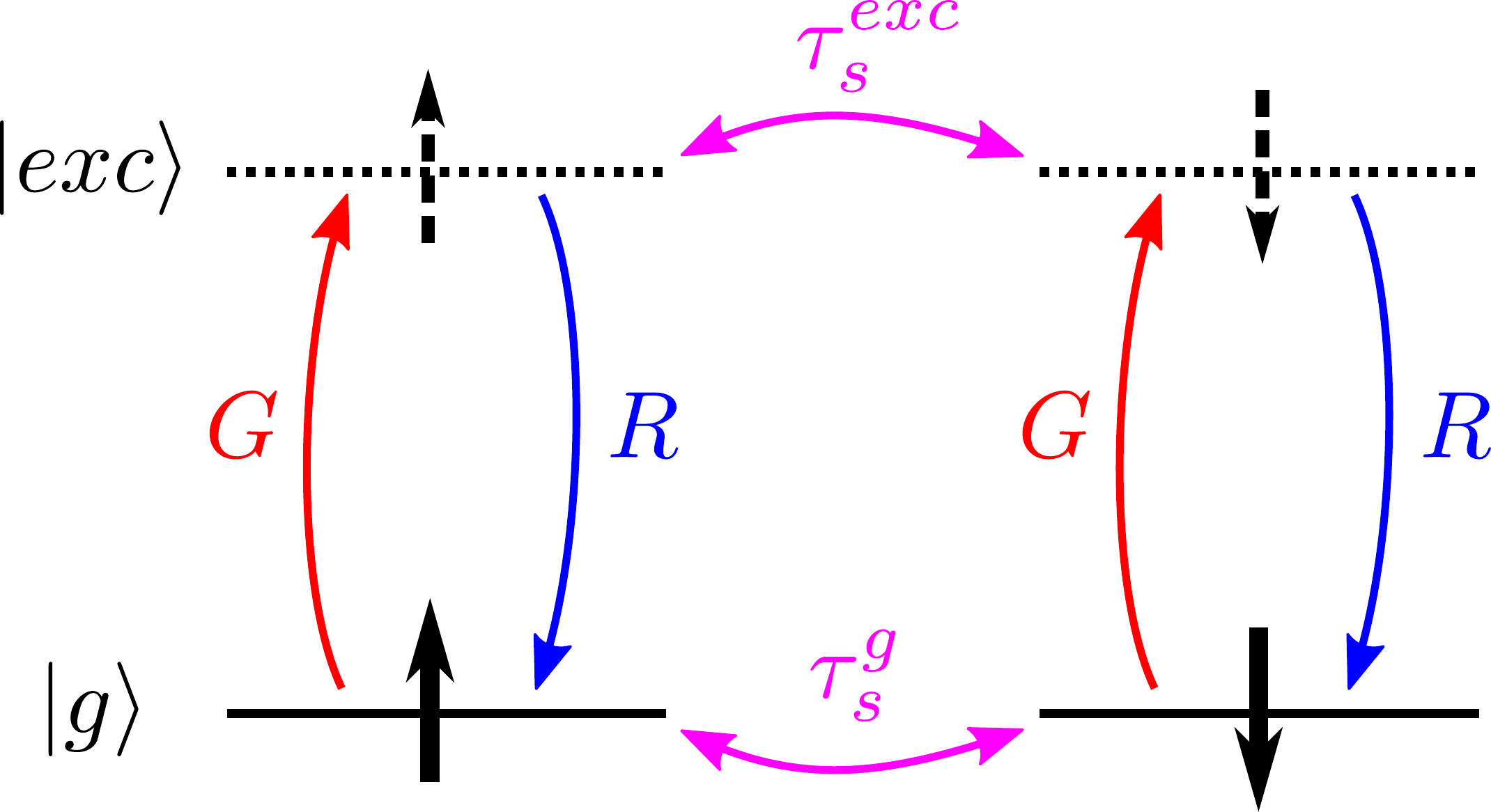}
\caption{\label{fig:4levels}Scheme of transitions between the ground, $|g\rangle$, and excited, $|exc\rangle$, states of the localization center and of spin relaxation in both states. Solid and dashed thick arrows denote the spin component, $S_z^{g,exc}$, in the ground and excited states, respectively. (Adapted from Ref.~\cite{Nonresonant_nonequilibrium}.)}}
\end{figure}

Experimentally, the most natural non-equilibrium situation arises when either the absorption of the probe beam or additional non-resonant excitation of the system creates nonequilibrium electrons and holes in the quantum dot. It affects both the dynamics of the system and microscopic mechanisms of the Faraday and Kerr rotation effects. A broad range of the non-equilibrium spin noise spectroscopy experiments can be described within the four-level model formulated in Refs.~[\onlinecite{glazov_keldysh}] and~[\onlinecite{noise-trions}]. In this model in addition to the spin-degenerate ground state a two-fold degenerate excited state, e.g., the trion state, is taken into account, see Fig.~\ref{fig:4levels}.

Let us assume that the excitation and recombination processes in the four-level model conserve the spin direction. The rates of these processes $G$ and $R$, respectively, do not depend on the spin orientation as shown in Fig.~\ref{fig:4levels}. Thus, the occupancy of the ground state $n_g$ satisfies the equation
\begin{equation}
  \dot{n}_g=-G n_g+R n_{exc},
\end{equation}
where $n_{exc}=1-n_g$ is the occupancy of the excited state. In the steady-state
\begin{equation}
  \label{eq:ng}
  n_g=\frac{R}{R+G}.
\end{equation}
Equations of motion for a spin in the ground state, $\bm{S}^g$, and excited state, $\bm{S}^{exc}$, have the form
\begin{subequations}
  \label{eq:4states}
  \begin{equation}
    \label{eq:4states_g}
    \frac{\d\bm S^g}{\d t}=\bm\Omega_g\times\bm S^g-\frac{\bm S^g}{\tau_s^g}-G\bm S^g+G\bm S^{exc}+\frac{1}{\tau_0}S_z^{exc}\bm e_z,
  \end{equation}
  \begin{equation}
    \label{eq:4states_exc}
    \frac{\d\bm S^{exc}}{\d t}=\bm\Omega_{exc}\times\bm S^{exc}-\frac{\bm S^{exc}}{\tau_s^{exc}}-R\bm S^{exc}+G\bm S^g,
  \end{equation}
\end{subequations}
where $\bm\Omega_{g,exc}$ are the frequencies of the spin precession in the ground and excited states due to both the external magnetic field and hyperfine interaction,  $\tau_s^{g,exc}$ are the phenomenological spin relaxation times, and $\bm e_z$ is a unit vector along the growth axis of the structure. We have taken into account in Eqs.~\eqref{eq:4states} that the rate of recombination (accounting for the two-level system saturation effects) can be recast as $R=G+1/\tau_0$, where $\tau_0$ is the spontaneous trion recombination time. In the latter process, according to the selection rules, only $z$-spin component is conserved~\cite{ast08,glazov_keldysh}. Spin noise spectra can be calculated analytically in this model solving the kinetic equations for the spin correlators.

\begin{figure}[t]
\centering
\includegraphics[width=\textwidth]{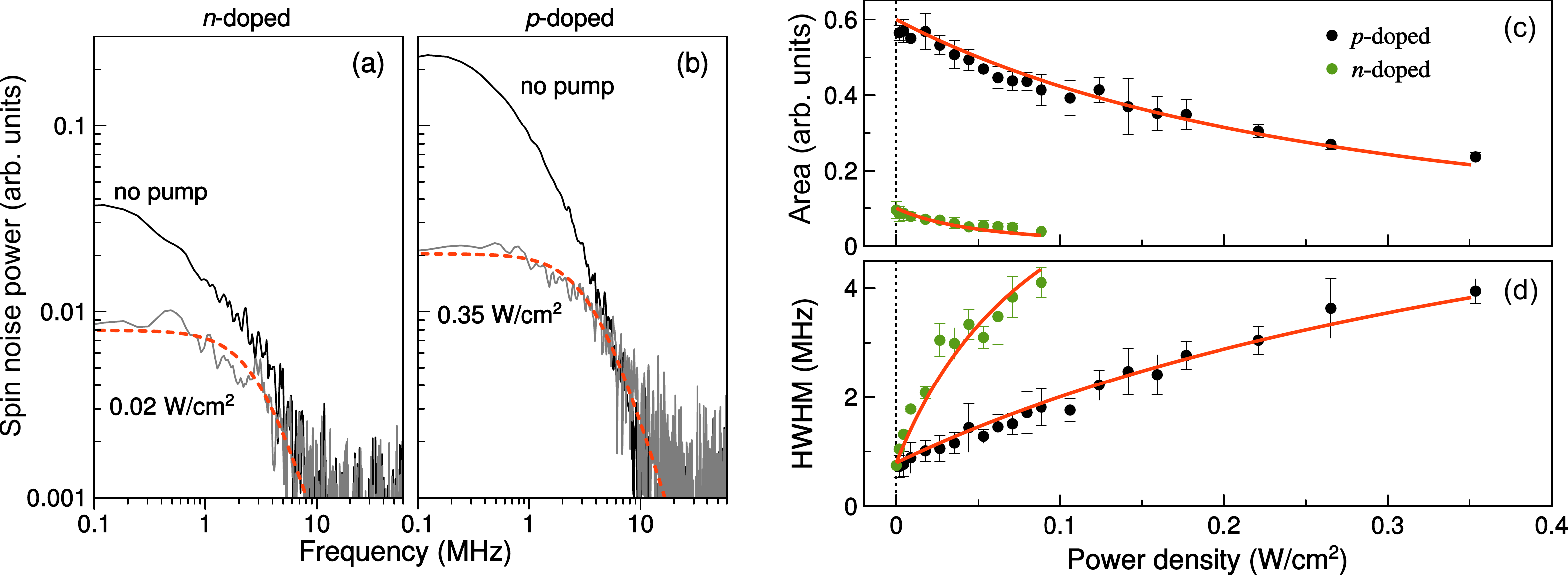}
\caption{\label{fig:nq}
Spin noise spectra in the absence and presence of the additional excitation, as indicated in  the figure measured for the samples containing (a)~$n$-type and (b)~$p$-type quantum dots. Red curves show Lorentzian fits. (c) Area under the spin noise spectrum as a function of the excitation power density  for $n$-type sample (green dots) and $p$-type (black dots). (d)~Analogous dependence of the half-width at the half-maximum. (Adapted from Ref.~\cite{Nonresonant_nonequilibrium}.)}
\end{figure}

The situation of a non-resonant excitation of the singly charged, by electrons or holes, quantum dots is the most illustrative one~\cite{Nonresonant_nonequilibrium}. Corresponding spin noise spectra are shown in Fig.~\ref{fig:nq}(a,b). The spin noise spectra have a Lorentzian form, whose area and half-width are shown in Fig.~\ref{fig:nq}(c,d) as functions of the excitation power density. The precessional peak in these spectra is at higher frequencies and not shown in the figure.

Despite the fact that under a non-resonant excitation there are many excited states instead of two, the model described above can quantitatively describe the spin noise spectra and their modification due to the generation of non-equilibrium electrons and holes.  Assuming that the generation and recombination processes are faster than the spin relaxation, one can show that the average rate of the spin relaxation is a weighted sum of the relaxation rates in the ground and excited states:
\begin{equation}
  \label{eq:tauss}
  \frac{1}{\tau_s^*}=\frac{n_g}{\tau_s^g}+\frac{n_{exc}}{\tau_s^{exc}},
\end{equation}
where the occupancies of the states are given by Eq.~\eqref{eq:ng}. The spin noise power is proportional to $n_g^2$. From the fit of the experimental data, one obtains the spin relaxation times $\tau_s^g=200$~ns and $\tau_s^{exc}=19$~ns for the both studied samples.

In order to enhance the spin signals, one can use microcavities which, however, also increase an amplitude of the electromagnetic field acting on the charge carriers~\cite{microcavities}. In a weak coupling regime the model described above can be used to calculate the non-equilibrium spin noise spectra of charge carriers localized at the quantum well width fluctuations taking into account the absorption of the probe beam and excitation of trions~\cite{noise-trions}. The spin relaxation of resident charge carriers becomes faster and anisotropic with increase of the excitation power and in the same time the effective spin precession frequency decreases. The reason for the anisotropy is related to the specifics of the selection rules, according to which the excitation of the trion and its consequent radiative recombination erases the spin components perpendicular to the growth axis of the structure, while this process does not affect the longitudinal spin component, see Eqs.~\eqref{eq:4states}.

\begin{figure}
  \centering
  \includegraphics[width=0.8\linewidth]{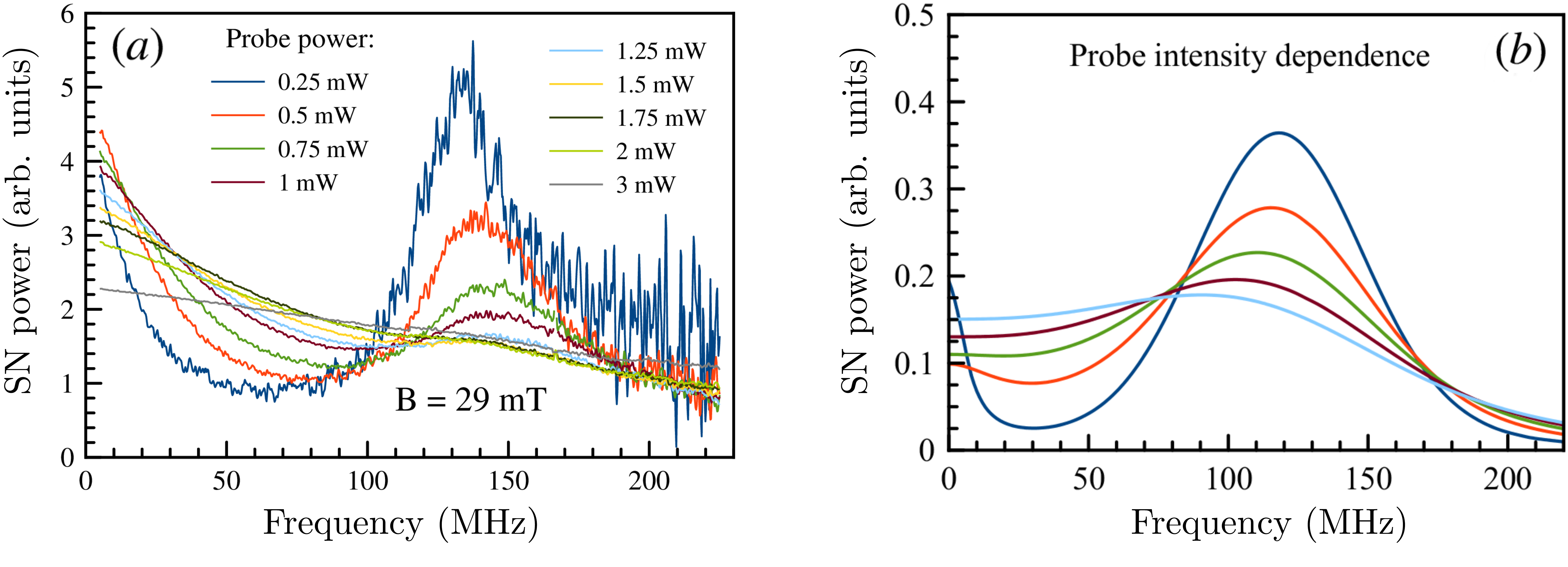}
  \caption{Spin noise spectra under the generation of singlet trions for the different optical excitation powers: (a) experiment, (b) theory. (Adapted from Ref.~\cite{noise-trions}.)}
  \label{fig:trions}
\end{figure}

Figure~\ref{fig:trions} shows the spin noise spectra measured experimentally [panel (a)] and calculated theoretically [panel (b)] for different intensities of the probe beam. Generally, the spectra consist of two peaks related to the spin relaxation (at the zero frequency) and the spin precession (at a positive frequency). An increase in the power of the beam results in the suppression of the precessional peak and enhancement of the relaxational peak.

Interestingly, the simplified description of the spin dynamics at the high power of the probe beam can be carried out in the terms of the quantum Zeno effect~\cite{khalfin1958contribution,Facchi_2008,Zeno_QED}. In agreement with general postulates of quantum mechanics, a continuous detection of the electron spin should result in the ``freezing'' of the spin dynamics due to the quantum back-action~\cite{doi:10.1063/1.523304}. It results in a renormalization of the spin precession frequency in the external transverse magnetic field   $\Omega_L$ as~\cite{Liu2010,Bednorz}
\begin{equation}
  \label{eq:omega_zeno}
  \Omega=\sqrt{\Omega_L^2-\lambda^2}.
\end{equation}
Here the phenomenological parameter  $\lambda$ characterizes the ``measurement strength'' which is proportional to the power of the probe beam. In the framework of the four-level model one can find that
\begin{equation}
  \lambda=\frac{G\tau_s^{exc}}{2(\tau_s^{exc}+\tau_0)},
\end{equation}
so both descriptions of non-equilibrium spin fluctuations are equivalent.

In a microcavity with a single quantum dot, a strong light-matter coupling regime can be realized~\cite{microcavities}. Electron spin polarization in such structures can result in macroscopic values of the polarization plane rotation angle~\cite{Arnold2015}, and it can also allow for a complete reflection of one of the circular components of the probe beam with a complete transmission of the other one. It is important to note, that for a single spin an ensemble averaging does not take place. Therefore, the fluctuations of the optical signals, particularly, that of the reflection and transmission coefficients of light passing through the system can be macroscopic as well.

A theoretical description of spin noise in the strong-coupling regime requires consideration of the quantum nature of the electromagnetic field. The correlation functions of the Faraday, Kerr and ellipticity signals can be expressed through the averages of the four field operators in the microcavity, in analogy with the second-order correlation function $g^{(2)}(\tau)$, which describes light intensity fluctuations~\cite{BackAction}. The latter depends on the electron spin dynamics due to the fact that, depending on the mutual orientation of the electron spin and photon angular momentum, the light can pass through the cavity or be reflected from it. The simplest optical signal reflecting the spin noise in such a system is the intensity transmission coefficient  $T$ of a circularly polarized light. Its correlation function, $\left\langle\delta T(t)\delta T(t+\tau)\right\rangle$, can be expressed as
\begin{equation}
  \frac{\left\langle\delta T(t)\delta T(t+\tau)\right\rangle}{T_0^2}=g^{(2)}(\tau)-1,
\end{equation}
with  $T_0$ being the average transmission coefficient.

Usually, the spin relaxation time and precession period in an external magnetic field are by far longer than the trion recombination time, the photon lifetime in the cavity, as well as the period of Rabi oscillations between the polariton states.  This separation of time-scales allows one to develop an analytical approach for calculation of the electron and photon dynamics in the system. In the framework of this approach, the expressions for the correlation function of the light transmission coefficients have been derived in the limit of small light intensities~\cite{BackAction}. These expressions agree well with the numerical calculations based on the density matrix formalism. The spectrum of the transmission coefficient noise, generally, consists of the spin and photon components which contain information, respectively, about the spin dynamics in the system and about the properties of the excited states, such as the polariton splitting and the lifetimes of the excited states.





\section{Spin chains}
\label{sec:1D}

Among the most intriguing objects for the analysis of spin dynamics and spin fluctuations are the one-dimensional spin chains. The interest in the spin diffusion and spin localization in chains is related, on the one hand, to the fact that the one-dimensional systems are traditionally used as objects for testing theoretical approaches that can be also applied to systems of larger dimensions, and, on the other hand, to the presence of a number of objects that demonstrate pronounced one-dimensional properties~\cite{Skalyo,Dingle,Birgeneau,Hutchings}. Here we will focus on one of the most studied and widely applied models of spin chains: spin-$1/2$ XXZ chain of $N$ sites with disorder described by the Hamiltonian
\begin{equation}
  H_{XXZ}=-J_{\perp}\sum_{m=1}^{N-1}(S_{m}^{x}S_{m+1}^{x}+S_{m}^{y}S_{m+1}^{y})-J_z\sum_{m=1}^{N-1}S_{m}^{z}S_{m+1}^{z}+\hbar\sum_{m=1}^N\Omega_mS_{m}^{z},
\end{equation}
where index $m$ numerates $N$ sites in the chain, $J_{\perp}$ and $J_{z}$ are the exchange interaction constants between the nearest-neighbor spins, $S_{m}^{\alpha}$ are spins components of the sites $m$ ($\alpha=x,y,z$) and $\Omega_{m}$ are the random magnetic fields at the chain sites oriented along the $z$ axis. Using a Jordan-Wigner transformation~\cite{J-W} one can map spin-$1/2$ XXZ chain model onto the chain of spinless fermions. In this mapping, exchange interaction constants $J_{\perp}$ and $J_{z}$ describe fermion hopping between nearest-neighbor sites and nearest-neighbor interaction, respectively, while the terms $\Omega_m S_m^z$ correspond to the random on-site energies. In this model, the projection of the total spin $\bm S=\sum_{m=1}^N \bm S_m$ on the $z$ axis is conserved.

In particular, for the XX model without disorder, when $J_{\perp}=J$, $J_{z}=0$ and $\Omega_m^z=0$ the Hamiltonian of the system after Jordan-Wigner transformation has a very simple form
\begin{equation}
  \label{eq:Hm}
  \mathcal H = \frac{J}{2}\sum_{m=1}^{N-1}\left( a_m^\dag a_{m+1} + {\rm{h.c.}}\right),
\end{equation}
where operators $a_m$ and $a_m^\dag$ obey the standard commutation rules for fermions and describe annihilation and creation of quasi-particles at the chain sites. The $z$ components of spins are given by
\begin{equation}
  S_m^z=a_m^\dag a_m-1/2,
\end{equation}
so the problem of spin fluctuations is reduced to the calculation of the correlation function of particles density. The Hamiltonian~\eqref{eq:Hm} is diagonalized by the Fourier transformation:
\begin{equation}
\label{Ham_spin_chains_simple}
\mathcal H=\sum_{k}E_{k}a_{k}^{+}a_{k}
\end{equation}
with the corresponding dispersion law
\begin{equation}
  \label{eq:cos_disp}
  E_{k}=J\cos(k),
\end{equation}
where $a_{k}=\sqrt{2/(N+1)}\sum_{m=1}^{N}\sin(km)a_m$ with $k=\pi n/(N+1)$, and index $n$ takes values $1,2,3,\ldots,N$.

An exact expressions for the correlation functions $\langle S_{m}^{z}(\tau)S_{m'}^{z}(0)\rangle$ between the chain sites with indexes $m$ and $m'$
were found for the XX model without disorder \cite{Niemeijer,Katsura}. In the limit of infinite number of sites in the chain, the expression for correlation functions of spin $z$-components has a simple form
\begin{equation}
\langle S_{m}^{z}(\tau)S_{m'}^{z}(0)\rangle=\frac{1}{4}J_{|\Delta m|}^{2}(J\tau/\hbar),
\end{equation}
where the order of the Bessel function $J_{|\Delta m|}(J\tau/\hbar)$ is determined by the modulus of the difference of chain sites numbers $\Delta m=m-m'$ and the argument of the Bessel function depends on the strength of the exchange interaction. Here, as it was done earlier, we assume that temperature exceeds by far all other energy scales in the system, but the thermalization time is long enough to neglect inelastic processes. XX model also allows one to obtain exact expressions for the transverse correlation functions $\langle S_{m}^{\alpha}(\tau)S_{m'}^{\alpha}(0)\rangle$, where $\alpha=x,y$, which correspond to the correlations of fermionic operators $a^\dag_m$, $a_{m'}$~\cite{Sur,Brandt,Capel,Florencio}. In the case of large number of sites and high temperatures spin correlation functions for $x$- and $y$-components are equal to each other and can be written as
\begin{equation}
\langle S_{m}^{x}(\tau)S_{m'}^{x}(0)\rangle=\langle S_{m}^{y}(\tau)S_{m'}^{y}(0)\rangle= \frac{1}{4}\e^{-\left(\frac{J\tau}{2\hbar}\right)^2}\delta_{m,m'},
\end{equation}
which corresponds to the spectrum of total spin noise
\begin{equation}
(S_x^2)(\omega)=\frac{N\hbar\sqrt{\pi}}{2J}\e^{-\left(\hbar\omega/J\right)^2}.
\end{equation}
These expressions demonstrate that spin fluctuations along $x$ and $y$ axes are localized and do not propagate, while spin fluctuations along $z$ axis propagate with the typical velocity $\propto J$~\cite{Luther,Fogedby}, i.e. non-diffusively. It was shown in Ref.~\cite{Bohm} that for the XXZ model with $J_{z}\neq0$ the behavior of correlation functions changes and can become diffusive.

Let us now analyze spin correlators in the presence of random fields $\Omega_m$, which can be caused, for example, by the hyperfine interaction with the nuclear spins of the crystal lattice. Following Eq.~\eqref{eq:F_e} let us consider the Gaussian form of the distribution function of random fields:
\begin{equation}
  \mathcal F(\Omega)=\frac{1}{\sqrt{\pi}\delta}\e^{-\Omega^2/\delta^2}.
\end{equation}
For the XX model, any small disorder leads to the localization of spin fluctuations. Formally, this problem is equivalent to the problem of non-interacting electrons localization in the 1D system with <<diagonal>> disorder~\cite{Lifshitz_book}.

Firstly, random fields cause the modification of density of states, which in the absence of disorder reads
\begin{equation}
  \rho_0(E)=\frac{1}{\pi\sqrt{J^{2}-E^{2}}}
\end{equation}
and diverges in the vicinity of $E=\pm J$, as velocity of carriers is equal to zero for these energy values, see Eq.~\eqref{eq:cos_disp}. As it was shown in Refs.~\cite{Halperin1965,Lifshitz_book}, weak disorder ($\delta\ll J$) modifies density of states $\rho$ in the vicinity of the band edges:
\begin{equation}
  \rho(E)=\frac{1}{\pi^2}\sqrt{\frac{E_0}{J}}\frac{\d}{\d\Delta E}\frac{1}{{\rm Ai}^2(-2\Delta E/E_0)+{\rm Bi}^2(-2\Delta E/E_0)},
\end{equation}
where $\Delta E=\pm E- J$, $E_0=\hbar\delta \sqrt[3]{\hbar\delta/J}$ and ${\rm Ai}(x)$, ${\rm Bi}(x)$ are Airy functions of the first and second kind, respectively.

Secondly, the presence of disorder leads to the localization of wave functions. The distribution function of the inverse localization length $w(A)$ can be found by averaging the distribution function of the inverse localization length for a given velocity over all the wave vectors \cite{Berezinskii,Gorkov}. The result reads:
\begin{equation}
\label{w_A}
  \langle w(A)\rangle=\frac{1}{2\pi\i}\int\limits_{-\i\infty}^{\i\infty}\frac{d xx}{A_0\sinh^2\left(\sqrt{x}\right)}\exp\left(\frac{x A}{A_0}\right)\left[I_0\left(\frac{xA}{A_0}\right)+I_1\left(\frac{xA}{A_0}\right)\right],
\end{equation}
where $A_0=3(\hbar\delta)^2/(2J^2)$. Integration is performed along the imaginary axis, $\Re(x)=0$.

\begin{figure}
\includegraphics[width=0.6\textwidth]{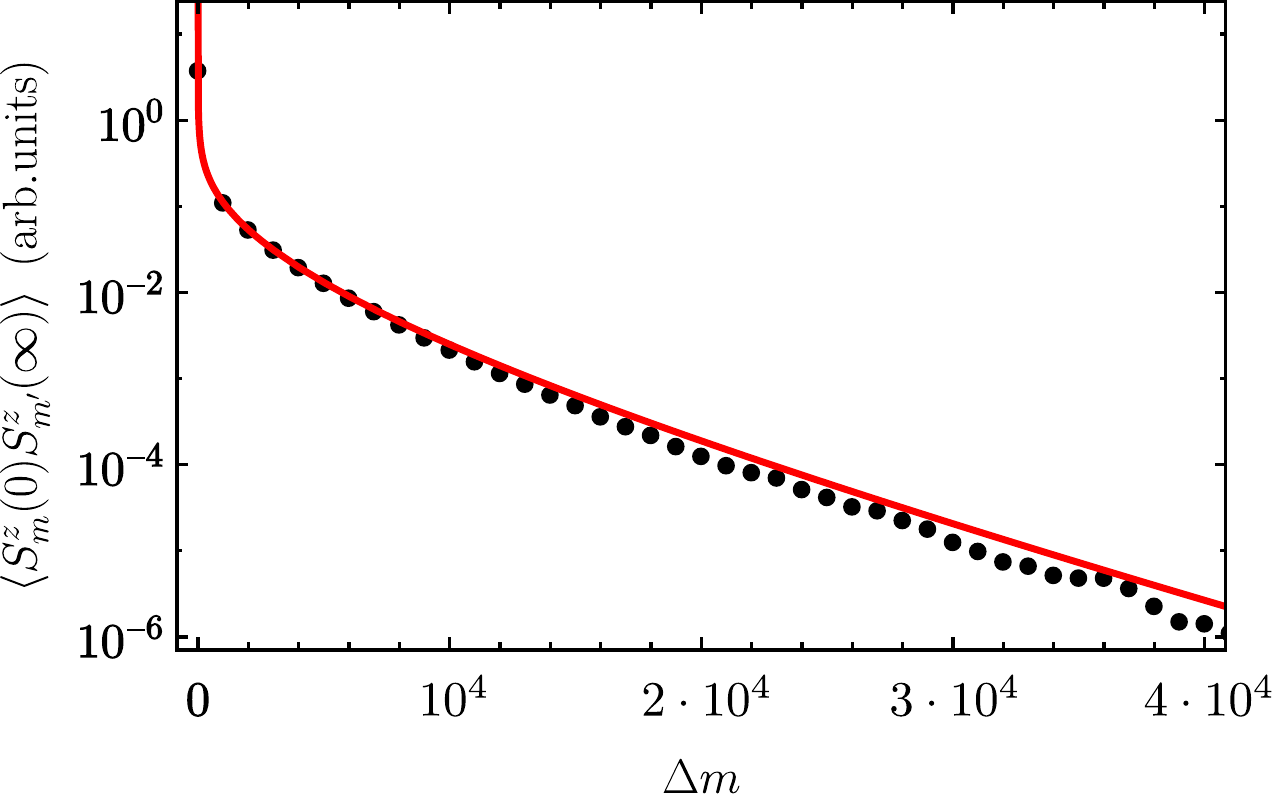}
\caption{Spin correlation function for the XX model with disorder in the long time limit. Solid curve is calculated analytically, dots correspond to the numerical calculations performed for the $10^5$ sites with periodical bound conditions for $\hbar\delta = 0.035J$.}
\label{z_corr}
\end{figure}

Similarly, one can find spin correlation function $\langle S_m^z(0) S_{m'}^z(\infty)\rangle$ in the long time limit, because in this limit one can neglect the interference effects between the wave functions corresponding to the different eigenstates. Figure~\ref{z_corr} shows the result of the numerical calculations performed for the $10^5$ sites. For a fixed value of the particle wave vector $k$, the localization length is equal to  $l_0=2J^2/(\hbar\delta)^2\sin^2(k)$~\cite{Berezinskii1974} and the explicit expression for the correlation function of particles density was found in Ref.~\cite{Gogolin1976}. Its long distances asymptotics has the form~\cite{Gogolin1975}:
\begin{equation}
  p(\Delta m)\approx\frac{\pi^3\sqrt{\pi}}{16l_0}\left(\frac{l_0}{\Delta m}\right)^{3/2}e^{-\Delta m/l_0}.
\end{equation}
Averaging of the exact correlation function over the wave vectors gives the possibility to obtain the expression for the spin correlation function in the limit of long distances and high temperature:
\begin{equation}
\label{eq:z_dis}
  \langle S_m^z(0) S_{m'}^z(\infty)\rangle=\frac{\pi^3J^2}{32\delta^2\Delta m^2}\exp\left[-\frac{\delta^2}{8J^2}\Delta m\right].
\end{equation}

For the XXZ model, the localization takes place in the presence of arbitrary small disorder. In this case a threshold of many-body localization can be studied~\cite{Prosen,Pal,MBL_Das_Sarma}. However, for the isotropic XXX model ($J_\perp=J_z$) the conservation of the total angular momentum makes complete many-body localization impossible~\cite{Parameswaran_2018}.

The role of inelastic processes in the dynamics of spin correlators in the 1D systems is still not well understood. The interplay between electrons hopping between the localized states and hyperfine interaction with nuclear spins of the lattice in the quantum wire was analyzed in Ref.~\cite{Sherman_QW}. In the case of low density of sites, the localization leads to the exponentially broad distribution of the hopping times. It leads to the low frequency singularity in the noise spectra. The dynamics of spin fluctuations in this regime can be well described by the model of pairs of closely localized sites and single states, similarly to the model of clusters, which was discussed in the previous section.


\section{Two dimensional systems}
\label{sec:2D}



The key difference of the two dimensional systems from the one dimensional systems with disorder and zero dimensional systems from the point of view of the spin dynamics and spin noise is the free propagation of the charge carriers. This leads to the considerable suppression of the hyperfine interaction, so that the spin dynamics is mainly driven by the spin-orbit interaction.

In the structures without inversion center, the effective Hamiltonian consists of the term responsible for the kinetic energy of a particle $\varepsilon_k=\hbar^2 k^2/2m$, where $\bm k$ is the quasiwavevector and $m$ is the effective mass of an electron, and the spin-dependent terms, which can be written in the form~\cite{dresselhaus55,rashba1959-2,dyakonov_book}
\begin{equation}
\label{H:so}
\mathcal H_{so} = \frac{\hbar}{2} (\bm \Omega_{\bm k} \cdot \bm \sigma).
\end{equation}
Here $\bm \Omega_{\bm k}$ is the pseudovector with the absolute value and direction strictly related to the electron wavevector $\bm k$. In the quantum well structures based on III-V and II-VI materials and in a number of systems based on silicon and germanium, the components of the vector $\bm \Omega_{\bm k}$ are linearly related with the components of $\bm k$. In particular, in quantum wells with the zinc blend structure grown along $z\parallel [001]$ direction, the pseudovector $\bm \Omega_{\bm k}$ can be presented in the following form~\cite{bychkov84,dyakonov:110,averkiev:15582,golub_ganichev_BIASIA}
\begin{equation}
\label{QW:001}
\bm\Omega_{\bm k} = (\beta_1 k_y, \beta_2 k_x,0),
\end{equation}
where the axes $x\parallel [1\bar 10]$ and $y\parallel [110]$ are the main axes of the $C_{2v}$ point symmetry group, which generally characterizes the symmetry of such structures. The constants $\beta_{1,2}$ are determined by the specific structure and microscopic mechanisms of the lifting of the spectrum spin degeneracy. In particular, the case of $\beta_1 = - \beta_2$ corresponds to the dominant bulk inversion asymmetry (the Rashba contribution to the spin splitting), and $\beta_1 = \beta_2$ corresponds to the bulk (or in some cases interface) inversion asymmetry (the Dresselhaus contribution).

It follows from the form of the effective Hamiltonian~\eqref{H:so} that the electron spin precesses in the effective magnetic field with the frequency $\bm\Omega_{\bm k}$, which determines the dynamics and relaxation of the spin fluctuations. The regime of the spin dynamics is determined by the relation between the typical frequency $\Omega_{k}$ and the electron collision time $\tau^*$ (strictly speaking, the single electron momentum relaxation time, which accounts for the scattering of the static disorder and phonons as well as electron-electron collisions~\cite{Glazov02r,leyland06}). Hereafter we assume, that the typical spin splitting $\hbar\Omega_k$ is much smaller than the electron kinetic energy, which allows us to neglect the effect of the spin dynamics on the orbital electron motion. In the limit of frequent collisions
\begin{subequations}
\label{frequent}
\begin{equation}
\label{cond:frequent}
\Omega_k \tau^* \ll 1,
\end{equation}
the relaxation of spin fluctuations is described by the exponential law and the correlation functions obey the system of equations (cf. Eq.~\eqref{eq:corr:gen} and Refs.~\cite{ll5,averkiev:15582})
\begin{equation}
\label{eqs:frequent}
\frac{d \mathcal C_{\alpha\beta}(\tau)}{d\tau} + \Gamma_{\alpha\alpha'}  \mathcal C_{\alpha'\beta}(\tau) = 0, \quad \Gamma_{\alpha\beta} = \langle(\Omega_{\bm k}^2 \delta_{\alpha\beta} - \Omega_{\bm k, \alpha} \Omega_{\bm k,\beta})\tau^* \rangle.
\end{equation}
In the tensor of the inverse spin relaxation times $\Gamma_{\alpha\beta}$ the averaging is performed over the electron ensemble described by the equilibrium Fermi-Dirac distribution function $f(\varepsilon_k)$ according to
\[
\langle F(\bm k) \rangle = \frac{\sum_{\bm k} F(\bm k) f'(\varepsilon_k)}{\sum_{\bm k}  f'(\varepsilon_k)}.
\]
\end{subequations}

The situation turns out to be fundamentally different when the collisions are relatively rare
\begin{subequations}
\label{rare}
\begin{equation}
\label{cond:rare}
\Omega_k \tau^* \gtrsim 1.
\end{equation}
In this case the electron spin rotates by a considerable angle between the consequent collisions, so the spin noise demonstrates the oscillating behaviour in time, which is analogous to the spin polarization oscillations after the pulsed optical excitation~\cite{gridnev01_ru,brand02,glazov2007,PhysRevB.80.241314}. As an example, let us consider the equations, that describe the dynamics of the correlation function of the $z$ component of the total spin of electron gas for arbitrary $\Omega_k\tau^*$. Let us introduce the auxiliary function
\[
C_{zz}(\tau;\bm k,\bm k') = \langle \{\delta s_{\bm k, z}(t+\tau) \delta s_{\bm k', z}(t)\}\rangle,
\]
where $s_{\bm k, z}(t)$ is the component of the spin polarization with the wave vector $\bm k$. One has $C_{zz}(\tau) = \sum_{\bm k,\bm k'} C_{zz}(\tau;\bm k,\bm k')$. This function obeys the equation
\begin{equation}
\label{eqs:rare}
\left(\frac{d}{d\tau} + \frac{1}{\tau^*}\right) \frac{d}{d\tau} C_{zz}(\tau;\bm k,\bm k') + \Omega_{\bm k}^2 C_{zz}(\tau;\bm k,\bm k') + \left(\frac{d}{d\tau} + \frac{1}{\tau^*}\right)\frac{C_{zz}(\tau;\bm k,\bm k') - \bar C_{zz}(\tau;\bm k,\bm k')}{\tau^*} =0,
\end{equation}
where the bar above the expression denotes averaging over the direction of the vector $\bm k$ for its fixed absolute value. This differential equation of the second order in $\tau$ is obtained from the equations of the first order for the correlator $\langle \{\delta s_{\bm k, z}(t+\tau) \delta s_{\bm k', z}(t)\}\rangle$ and the combination $\Omega_{\bm k,y}\langle \{\delta s_{\bm k, x}(t+\tau) \delta s_{\bm k', z}(t)\}\rangle-\Omega_{\bm k,x}\langle \{\delta s_{\bm k, y}(t+\tau) \delta s_{\bm k', z}(t)\}\rangle$ accounting for the fact that the frequency $\bm\Omega_{\bm k}$ is perpendicular to the $z$ axis. In particular, for the degenerate electrons, isotropic spin splitting in the quantum well plane and $\Omega_{k_F} \tau^* \gg 1$, where $k_F$ is the Fermi wave vector, one has $C_{zz}(\tau) \propto \cos{(\Omega_{k_F} t)}\exp{(-t/2\tau^*)}$~\cite{gridnev01_ru}.
\end{subequations}

Interestingly, the degeneracy of the electron gas leads to the suppression of the spin noise intensity. In can be easily seen from the direct calculation of the equilibrium values of the correlators
\begin{equation}
\label{corr:degen}
\mathcal C_{\alpha\beta}(0) = \frac{\delta_{\alpha\beta}}{4} \sum_{\bm k} f(\varepsilon_k)[1-f(\varepsilon_k)] \sim
\frac{N}{4}\begin{cases}
\frac{k_B T}{E_F}, \quad k_B T \ll E_F = \frac{\hbar^2k_F^2}{2m},\\
1, \quad k_B T \gg E_F.
\end{cases}
\end{equation}
Here $N$ is the electron concentration. As expected from the qualitative arguments, in the degenerate gas, the electron spins fluctuate only in the states in the energy window $k_B T$ in the vicinity of the Fermi energy. The filled states with $E_F-\varepsilon \gg k_BT$ do not contribute, since in each this state there are two electrons with the opposite spins. A similar suppression takes place also in the three dimensional systems with the free electrons and was experimentally observed in GaAs crystals~\cite{PhysRevB.79.035208}. In the finite area, $A$, even at zero temperature the quantum correlations~\cite{ll5} give rise to the additional contribution $\mathcal C_{\alpha\beta}(0)\propto k_F\sqrt{A}\ln(k_F^2A)$, which is related with the finite wavelength of the electrons.

The linear relation between the magnetic field $\bm\Omega_{\bm k}$ and the electron wave vector $\bm k$, as well as the possibility of the free electron propagation in the quantum well plane, lead to the number of the vivid features of spatio-temporal spin correlators. Let us illustrate these effects by the case of the compensation of the Dresselhaus and Rashba terms $\beta_1 \ne 0$, $\beta_2 =0$. In this case, the spin precession axis coincides with the structure $x$ axis and the spin rotation angle during the electron motion is determined only by its displacement along the $y$ axis, $\Delta y$:
\begin{equation}
\label{helix:angle}
\theta = \frac{m \beta_1}{\hbar} \Delta y.
\end{equation}
Notably, in this situation the spin fluctuations in the different places are correlated or anticorrelated depending on the value of $\theta$ in Eq.~\eqref{helix:angle} and independently of the ballistic or diffusive electron motion. In this situation the persistent spin helix emerges~\cite{gridnev02,Schliemann03,Bernevig06,Koralek2009}. If $\beta_1=0$, $\beta_2 \ne 0$, the spin helix forms not along $y$ axis, but along $x$ axis. The absence of the exact compensation of the structure and bulk asymmetries in the spin splitting as well as the presence of terms cubic in the wave vector in the effective Hamiltonian lead to the violation of the relation~\eqref{helix:angle}, but in the broad range of parameters the long living spin modes with nontrivial spin correlations are formed in the two dimensional structures.

The consistent theory of spin noise with spatial and time resolution is developed in Ref.~\cite{poshakinskiynoise}, see also a review~\cite{Passmann_2019}, where the experimental data concerning the observation of the spin helix in the spin diffusion are given. Spatio-temporal spin noise is characterized by the correlation function [cf. Eq.~\eqref{correlations:qnt}]
\begin{equation}
\label{space:time}
\mathcal K_{\alpha\beta}(\bm \rho, \tau) = \langle\{\delta s_{\alpha}(\bm r+\bm \rho,t+\tau) \delta s_{\beta}(\bm r,t)\}_s\rangle,
\end{equation}
where the averaging is performed over $t$ and $\bm r$ (with the fixed $\tau$ and $\bm \rho$). In the case of frequent collisions~\eqref{cond:rare}, the authors of Ref.~\cite{poshakinskiynoise} derived the compact analytical expression for the Fourier transform of the correlator~\eqref{space:time}:
\begin{equation}
\label{space:time:q:omega}
\mathcal K_{\alpha\beta}(\bm q, \omega) = \iint \mathcal K_{\alpha\beta}(\bm \rho,\tau)e^{\mathrm i \omega \tau - \mathrm i \bm q \bm \rho} d\bm \rho d\tau=\frac{m k_B T}{16\pi \hbar^2} \left(\frac{1}{-\mathrm i \omega + \hat{\Gamma} + Dq^2 + \mathrm i2\tau^*\hat{\Lambda}({\bm q})} + {\rm H.c.} \right)_{\alpha\beta}.
\end{equation}
Here $D$ is the electron diffusion coefficient, $\hat{\Gamma}$ is the tensor of inverse spin relaxation times~\eqref{eqs:frequent}, and the tensor $\hat{\Lambda}$ has the components
\[
\Lambda_{\alpha\beta}(\bm q) = -\epsilon_{\alpha\beta\gamma} \hbar\int_0^{2\pi} \Omega_{\gamma,\bm k} (\bm q\cdot \bm k) \frac{d\varphi_{\bm k}}{2\pi m^*},
\]
where $\varphi_{\bm k}$ is the polar angle of the vector $\bm k$. This tensor describes the electron spin precession and diffusion. The electron gas is assumed to be degenerate.

\begin{figure}
  \centering
  {\includegraphics[width=0.5\linewidth]{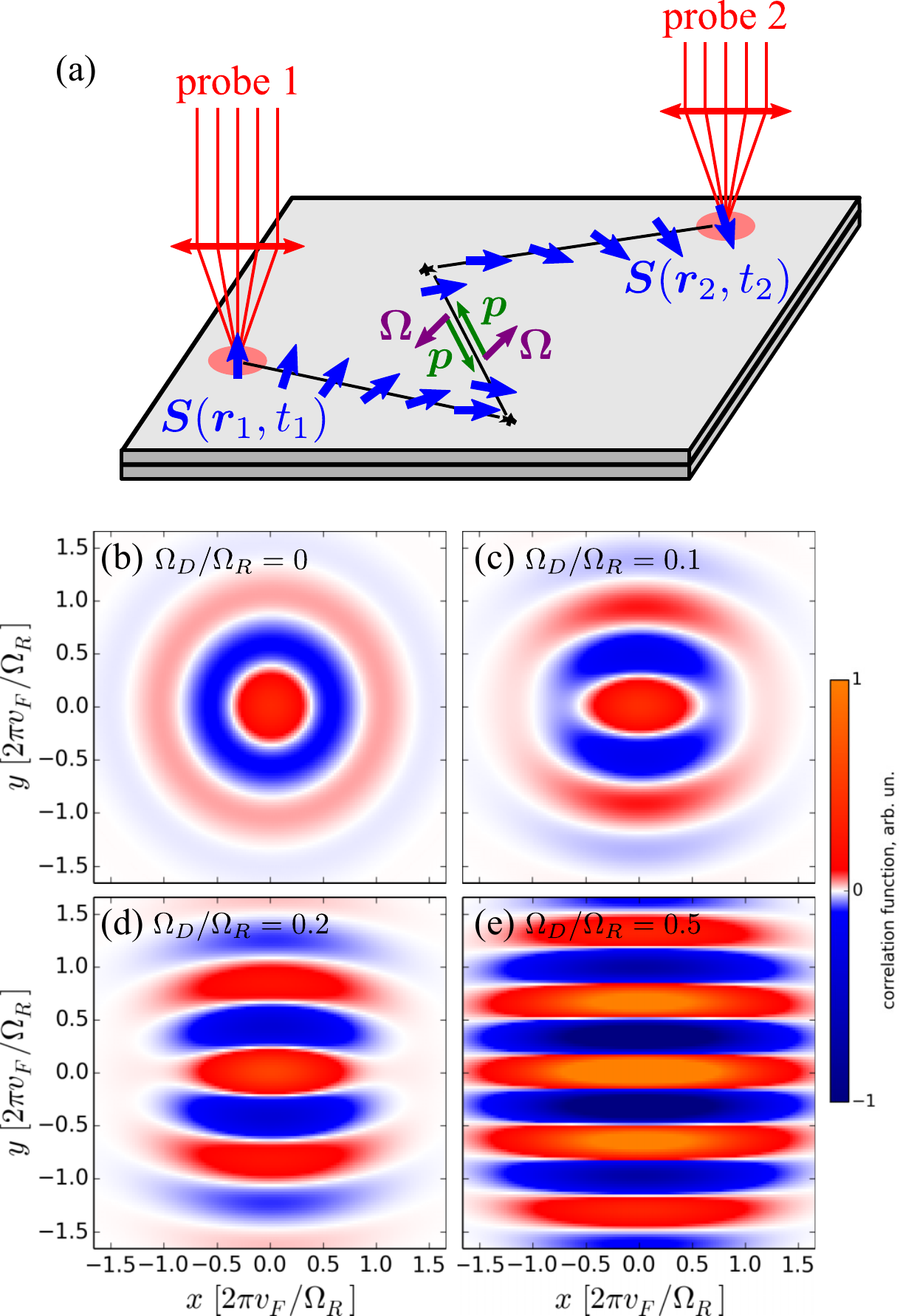}}
  \caption{\label{fig:SS} (a) Sketch of the possible experimental measurement of spatial correlation functions of the spin density fluctuations in the quantum well. (b)-(e) Spatial dependence of the correlator of the normal component of the spin density $\mathcal K_{zz}(\bm \rho,\tau)$ in the quantum well. The effective Rashba and Dresselhaus fields are introduced according to $\Omega_R=(\beta_1-\beta_2)k_F/2$, $\Omega_D=(\beta_1+\beta_2)k_F/2$. The calculation is performed for $\tau = 20/(\Omega_R^2\tau^*)$, $\Omega_R\tau^* = 0.2$ and different values of the Dresselhaus field. (Adapted from Ref.~\cite{poshakinskiynoise}.)}
\end{figure}

In Fig.~\ref{fig:SS}(a) a possible scheme is presented for the experimental detection of the spatio-temporal spin noise in the structure with a quantum well. Panels (b-e) show the results of the calculation of the spatial distribution of the spin density correlation functions performed in Ref.~\cite{poshakinskiynoise}. One can distinctly see the spatial oscillations of the spin polarization even in the cases, when the Rashba and Dresselhaus contributions to the spin splitting are considerably different. This is related with the fact that despite the many trajectories of the electron propagation in the diffusive regime between the given positions, the main contribution to the spin correlator is given by the trajectories close to the rectilinear one.

Above the situation when the spin splitting $\bm\Omega_{\bm k}$ does not depend on the coordinates was considered. There is a number of systems, where the spin splitting of the bands is absent or it is significantly suppressed ``on average'', but the spatial fluctuations are still present, see a review~\cite{GlazovSherman_rev}. The spin dynamics in these systems can be considerably different from the case of coordinate independent spin splitting. One may expect, that the spin noise in these system will demonstrate a number of specific features as well. This question remains poorly investigated, but in the quasi one dimensional systems with the spatial fluctuations of the spin splitting the power law divergence at low frequencies is expected~\cite{PhysRevLett.107.156602}.

Now let us turn to the description of the spin noise in two dimensional systems out of thermal equilibrium. The role of the spin-orbit interaction is particularly important if an external electric field $\bm E$ is applied to the system and causes the electron drift. The nonzero average wave vector $\langle \bm k_{dr} \rangle = (m\mu/\hbar) \bm E$, with $\mu$ being the mobility of the charge carriers, leads due to the spin-orbit interaction to the nonzero average effective magnetic field, which causes the regular precession of the spin fluctuations with the frequency $\bm \Omega_{\bm k_{dr}}$. This leads to the current induced shift of the peak in the spin noise spectrum. This effect was proposed and theoretically described in Ref.~\cite{Sinitsyn}. If the external magnetic field is absent, the peak in the spin noise spectrum is centered at the frequency $\Omega_{\bm k_{dr}}$, or at the frequency $|\bm \Omega_{\bm k_{dr}}+\bm \Omega_L|$, with $\bm \Omega_L$ being the Larmor spin precession frequency, if magnetic field is present. Experimentally, an analogous effect was observed in the electron spin resonance measurements in noncentrosymmetric SiGe quantum wells~\cite{wilamowski07}.

With increase of the electric current, the nonequilibrium effects in the spin noise spectra become even more pronounced~\cite{springerlink:10.1007}. The general theory of the spin noise in such conditions is described in Ref.~\onlinecite{glazov_keldysh}.

In Ref.~\onlinecite{NoiseStreaming}, a microscopic theory of spin dynamics and fluctuations of electron gas is developed for the streaming regime accounting for the spin-orbit interaction. The streaming regime is realized in relatively clean semiconductor structures under the application of moderately strong electric fields. It is characterized by the ballistic electron acceleration to the optical phonon energy during the time $t_{\rm tr}$, followed by the emission of phonon, loss of energy and return back to the region of small energies~\cite{Firsov_book}.

In Ref.~\onlinecite{NoiseStreaming} the kinetic equation is derived, that describes the spin dynamics in the streaming regime accounting for the spin-orbit interaction. It turns out to be convenient to separate the region in the momentum space with the small momentum component in the direction perpendicular to the applied electric field (the so-called ``needle'') from the rest of the momentum space. The dominant part of electrons in the streaming regime are in the ``needle''. The elastic scattering on impurities or quasielastic scattering on acoustic phonons lead to the scattering of electrons from the needle to the rest of the momentum space. The separation of the two contributions to the spin distribution function allows one to reduce the description of the spin dynamics to the determination of the spin noise spectrum and analysis of the eigenmodes in the system.

The damping rates of the different spin modes can be considerably different. Depending on the relation between Rashba and Dresselhaus spin splittings, the spin distribution in the momentum space can quickly relax to the homogeneous one (zeroes eigenmode) or to the oscillating one (one of the higher modes). In the latter case, the spin helix in the momentum space arises. The damping of the most long living spin mode can be determined by the combination of the spin-orbit coupling and the quasielastic scattering analogous to the Dyakonov-Perel mechanism~\cite{DP_rus}, or by the electron penetration in the region, where the energy is larger than the optical phonon energy, depending on the system parameters.

\begin{figure}
  \centering
  {\includegraphics[width=0.5\linewidth]{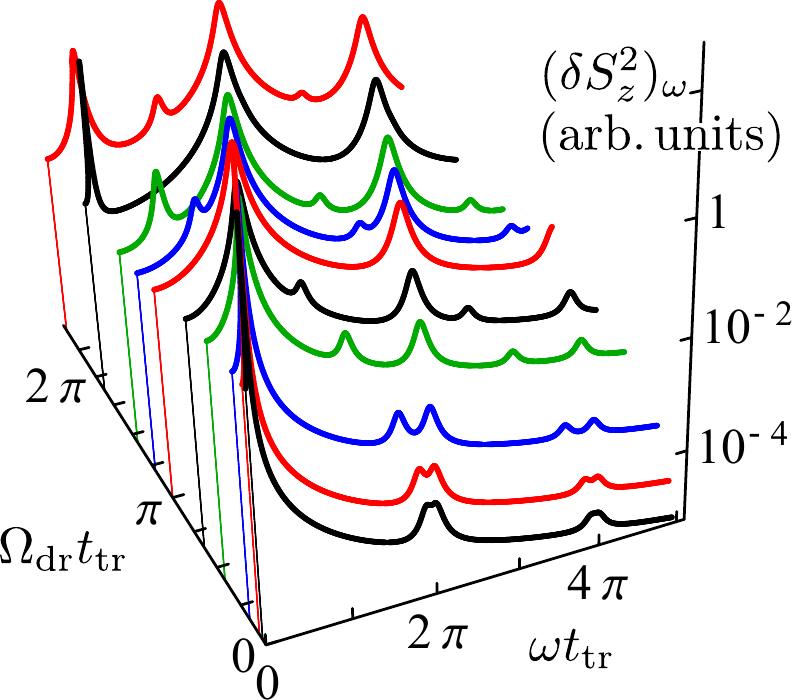}}
  \caption{\label{fig:streaming}Electron spin noise spectra in the streaming regime for the different values of the external electric field. (Adapted from Ref.~\onlinecite{NoiseStreaming}.)}
\end{figure}

The complex spin dynamics in the system manifests itself in the spin noise spectra. The noise spectrum consists of a series of peaks with the positions determined by the time of acceleration to the optical phonon energy $t_{\rm tr}\propto 1/E$ and the average spin precession frequency in the spin-orbit field, $\Omega_{\rm dr}$. The shape of the spectra for the different values of the parameter $\Omega_{\rm dr}t_{\rm tr}$ is shown in Fig.~\ref{fig:streaming}. The central frequencies and widths of the peaks are determined by the eigenfrequencies and decay times of the corresponding spin modes in the system. The highest peak is centered at the frequency $\Omega_{\rm dr}$. However, in the case of $\Omega_{\rm dr}t_{\rm tr}= 2\pi k$, where $k$ is an integer number, the spin helix in the momentum space emerges in the system, as mentioned above, and the electron spin rotates by an angle that is a multiple of $2\pi$ between the two consequent phonon emissions independently of the number of elastic scatterings during this time. In this case the amplitude of the peak at the lowest frequency drastically increases, see the black curve corresponding to $\Omega_{\rm dr}t_{\rm tr}\approx 2\pi$.

Apart from the electron systems, the exciton and exciton polariton systems are actively studied from the spin noise viewpoint. The key feature of these systems is the fact that they are fundamentally nonequilibrium, because excitons in quantum wells and exciton polaritons in quantum microcavities have finite lifetime, so the considerable role in these systems is played by the pumping and the fluctuations of number of quasiparticles accompanying it (generation-recombination noise). Moreover, excitonic systems have a rich fine structure: in the quantum wells, there are optically active (bright) excitons with the projection of the total angular momentum of the electron hole pair on the growth axis $m_z=\pm 1$, as well as the dark states with $m_z=\pm 2$. In Ref.~\onlinecite{PhysRevB.90.085303} the spin noise theory of bright and dark excitons is developed for the quantum wells in transverse magnetic field, and the interplay between electron-hole interaction leading to the splitting between the states with $|m_z|=1$ and $|m_z|=2$ and the external magnetic field induced mixing of bright and dark states is analyzed. In Refs.~\onlinecite{glazov_sns_pol,PhysRevB.91.161307} the spin noise of exciton polaritons is studied theoretically. Here important effects are the deceleration of the noise due to the Bose stimulation effect, modification of the spin noise statistics with increase of the pump power~\cite{PhysRevB.93.241307}, and the considerable role of the particle-particle interactions~\cite{PhysRevB.91.161307}. It was shown theoretically and experimentally, that in the structures with the quantum microcavities the giant enhancement of the spin noise can take place due to the effects of optical instability~\cite{ryzhov:jap}.




\section{High order spin correlators}
\label{sec:high}

In the previous sections we considered the spin correlation functions of the second order only. These correlators carry information about spin dynamics, but generally do not describe it completely. The complete information about spin properties is contained in the full set of correlators of all orders. For the classical fluctuating quantity $\bm S(t)$ the $n$-th order correlator has the form
\begin{equation}
  \label{eq:corr_n}
  \braket{\delta S_{z}(t_1)\delta S_z(t_2)\ldots\delta S_z(t_{n})}.
\end{equation}
Generally, for quantum operator $\delta \hat{S}_z(t)$ this expression should be symmetrized~\cite{Bednorz}:
\begin{equation}
  \label{eq:corr_n_sym}
  \braket{\left\{\delta\hat{S}_{z}(t_1)\left\{\delta\hat{S}_z(t_2)\ldots\delta\hat{S}_z(t_{n})\right\}_s...\right\}_s},
\end{equation}
cf. Eq.~\eqref{correlations:qnt}. Here the order $t_1<t_2<\ldots<t_n$ is assumed. Here as above we consider the thermal energy much larger than the Zeeman splitting of spin states, so there is no average spin polarization, $\braket{S_z(t)}=0$, and all the correlators of odd order vanish. 

For an ensemble of $N$ independent spins it is convenient to study instead of the correlation function~\eqref{eq:corr_n} the cumulants~\cite{Kubo_Cumulant,Mendel,Gardiner}. They are defined by the generating function
\begin{equation}
  \label{eq:generation}
  \ln\braket{\exp\left(\sum_{i=1}^nx_i\delta S_z(t_i)\right)}.
\end{equation}
In particular, the analog of Eq.~\eqref{eq:corr_n} is the coefficient $C_n$ of $x_1x_2\ldots x_n$ in the decomposition of generating function into Taylor series. For example, for $n=2$ and $n=4$ the cumulants have the form
\begin{subequations}
  \begin{equation}
    C_2\left\{\delta S_z(t)\right\}=\braket{\delta S_z(t_1)\delta S_z(t_2)},
  \end{equation}
  \begin{multline}
    \label{eq:cum4}
    C_4\left\{\delta S_z(t)\right\}=\braket{\delta S_z(t_1)\delta S_z(t_2)\delta S_z(t_3)\delta S_z(t_4)}-\braket{\delta S_z(t_1)\delta S_z(t_2)}\braket{\delta S_z(t_3)\delta S_z(t_4)}\\-\braket{\delta S_z(t_1)\delta S_z(t_3)}\braket{\delta S_z(t_2)\delta S_z(t_4)}-\braket{\delta S_z(t_1)\delta S_z(t_4)}\braket{\delta S_z(t_2)\delta S_z(t_3)}.
  \end{multline}
\end{subequations}
Generally, in each average the products should be symmetrized similarly to Eq.~\eqref{eq:corr_n_sym}.

The advantage of the cumulants as compared with the usual correlators is the additivity, which can be seen from the generating function~\eqref{eq:generation}. For the total spin composed of $N$ independent contributions $\bm S_k(t)$
\begin{equation}
  \bm S(t)=\sum_{k=1}^N\bm S_k(t),
\end{equation}
the cumulants are the sums of independent contributions as well:
\begin{equation}
  C_n\left\{\delta S_z(t)\right\}=\sum_{k=1}^NC_n\left\{\delta S_{k,z}(t)\right\}.
\end{equation}
One can see, that for $N\gg 1$, the dominant contribution to the correlator~\eqref{eq:corr_n} is given by the second order cumulant, and the higher order cumulants can be neglected. Thus, the noise of many independent spins is Gaussian, which means that the high order correlators can be calculated using the Wick's theorem.

The noise of a few spins is, generally, non-Gaussian. For example, for a single electron spin ($S=1/2$) the same-time fourth order cumulant is not zero:
\begin{equation}
  C_4=\braket{\delta S_z^4}-3\braket{\delta S_z^2}^2=-1/8.
\end{equation}
For different times $t_1<t_2<t_3<t_4$ the cumulant is given by
\begin{equation}
  \label{eq:C412}
  C_4\left\{\delta S_z(t)\right\}=-\braket{\delta S_z(t_1)\delta S_z(t_3)}\braket{\delta S_z(t_2)\delta S_z(t_4)}-\braket{\delta S_z(t_1)\delta S_z(t_4)}\braket{\delta S_z(t_2)\delta S_z(t_3)}.
\end{equation}

\begin{figure}
  \centering
  \includegraphics[width=0.7\linewidth]{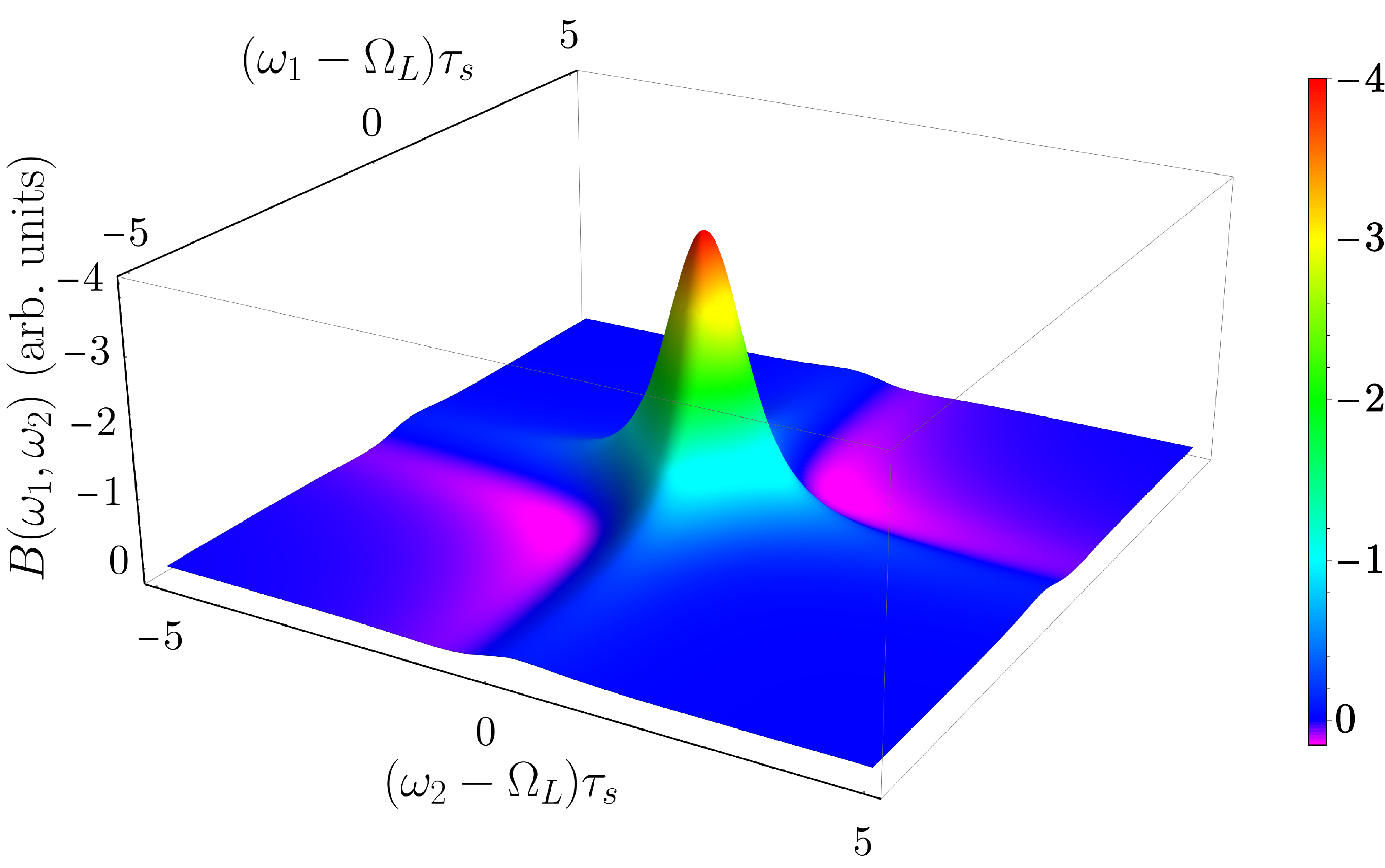}
  \caption{Spin noise bispectrum in the transverse magnetic field, calculated after Eq.~\eqref{eq:B4}. The vertical axis is directed downwards for better visibility, so the small positive values of $B(\omega_1,\omega_2)$ are shown by the magenta areas.}
  \label{fig:B4}
\end{figure}

The cumulant of the $n$-th order depends on $n-1$ time intervals, so its spectrum depends on $n-1$ frequencies. The simplest illustration of the spectrum of the high order is the bispectrum, which is the section of the fourth order spin noise spectrum. It is defined as a Fourier transform of the correlator~\eqref{eq:cum4} at the times $t_1=t$, $t_2=t+\tau_1$, $t_3=t+\tau$, $t_4=t+\tau+\tau_2$, integrated over $\tau$:
\begin{equation}
  B(\omega_1,\omega_2)=\iint\d\tau_1\d\tau_2\e^{\i\omega_1\tau_1+\i\omega_2\tau_2}C_4(\tau_1,\tau_2),
\end{equation}
where
\begin{equation}
  C_4(\tau_1,\tau_2)=\int\d\tau C_4\{\delta S_z(t)\delta S_z(t+\tau_1)\delta S_z(t+\tau)\delta S_z(t+\tau+\tau_2)\}.
\end{equation}
The bispectrum reflects the correlation degree of the spin noise at the frequencies $\omega_1$ and $\omega_2$. In the calculation of the bispectrum of the quantum noise, the operators should be symmetrized as in Eq.~\eqref{eq:corr_n_sym}. As a result, for ensemble of electrons using Eq.~\eqref{eq:C412} we obtain
\begin{multline}
  {NC_4(\tau_1,\tau_2)}=-(|\tau_1|+|\tau_2|)C_2(\tau_1)C_2(\tau_2)\\-\int\limits_{-\infty}^\infty\left[C_2(\tau-|\tau_2|)C_2(\tau+|\tau_1|)+C_2(|\tau|+|\tau_1|)C_2(|\tau|+|\tau_2|)\right]\d\tau,
\end{multline}
where
\begin{equation}
  C_2(\tau)=\braket{\{\delta S_z(t)\delta S_z(t+\tau)\}_s}.
\end{equation}
Provided the spin fluctuations precess in the transverse magnetic field with the frequency $\Omega_L$ and relax during the time $\tau_s$ [as in the derivation of Eq.~\eqref{noise:field}], the bispectrum at positive $\omega_1$ and $\omega_2$ has the form:
\begin{equation}
  \label{eq:B4}
  {NB(\omega_1,\omega_2)}=\frac{\tau_s^3{[(\delta_1+\delta_2)^2+4](\delta_1\delta_2-1)}}{16(1+\delta_1^2)^2(1+\delta_2^2)^2},
\end{equation}
where $\delta_{1,2}=(\omega_{1,2}-\Omega_L)\tau_s$ and it is assumed that $\Omega_L\tau_s\gg1$. This expression is shown in Fig.~\ref{fig:B4}. The bispectrum is centered at $\omega_1=\omega_2=\Omega_L$, and in the region where $\omega_1-\Omega_L$ and $\omega_2-\Omega_L$ have the same sign the bispectrum can be positive, and it is negative in the other region. These cases can be interpreted as positive and negative correlations between the noise at frequencies $\omega_1$ and $\omega_2$~\cite{sinitsynreview}.

The approach described above is valid for the description of the so-called ``weak'' quantum mechanical measurements, when the measurement hardly changes the density matrix of the system~\cite{Clerk}. The finite strength of the measurement can be described using the Krauss operators~\cite{kraus1983states,PhysRevLett.60.1351}
\begin{equation}
  K(s)=\left(\frac{2\lambda}{\pi}\right)^{1/4}\e^{-\lambda(s-S_z)^2},
\end{equation}
where $S_z$ is the spin operator, $s$ is continuous real parameter, and $\lambda$ describes the measurement strength. The probability of the spin measurement is $1-\e^{-\lambda/2}$. After a pulsed measurement the density matrix takes the form
\begin{equation}
  \rho(s)=K(s)\rho K(s),
\end{equation}
and it freely evolves between the measurements. After $n$ measurements, the average of the spin correlation function can be found as
\begin{equation}
  \label{eq:corr_Krauss}
  \braket{S_z(t_1)S_z(t_2)\ldots S_z(t_n)}=\iint\ldots\int s_1s_2\ldots s_n\Tr[\rho(s_1,s_2,\ldots,s_n)]\d s_1\d s_2\ldots\d s_n,
\end{equation}
where the parameters $s_i$ ($i=1,2,\ldots,n$) characterize the measurements at times $t_i$.

In the limit of weak measurements, when the parameter $\lambda\to0$ this definition of the correlation function coincides with Eq.~\eqref{eq:corr_n_sym}~\cite{Bednorz_PRL}. In the limit of the strong measurements ($\lambda\to\infty$) the correlator can be rewritten as
\begin{multline}
  \braket{S_z(t_1)S_z(t_2)\ldots S_z(t_n)}=\sum_{m_1,m_2,\ldots,m_n}m_1m_2\ldots m_n
  \\\times
  \Tr[P_{m_n}U_{t_n-t_{n-1}}(P_{m_{n-1}}U_{t_{n-1}-t_{n-2}}(\ldots U_{t_2-t_1}(P_{m_1}\rho P_{m_1})\ldots)P_{m_{n-1}})P_{m_n}],
\end{multline}
where $m$ are the eigenvalues of the operator $S_z$, $P_m$ are the projectors to the corresponding eigenstates ($S_z=\sum_mmP_m$), and superoperator $U_\tau(\rho)$ describes the free evolution of the density matrix during the time $\tau$. In fact, this expression describes the average over all possible trajectories of the system evolution between the spin values $m_i$ at the time moments $t_i$, and at each measurement the density matrix of the system is projected to the corresponding state (the eigenvalues $m_i$ are assumed to be nondegenerate).

\begin{figure}
  \centering
  \includegraphics[width=0.6\linewidth]{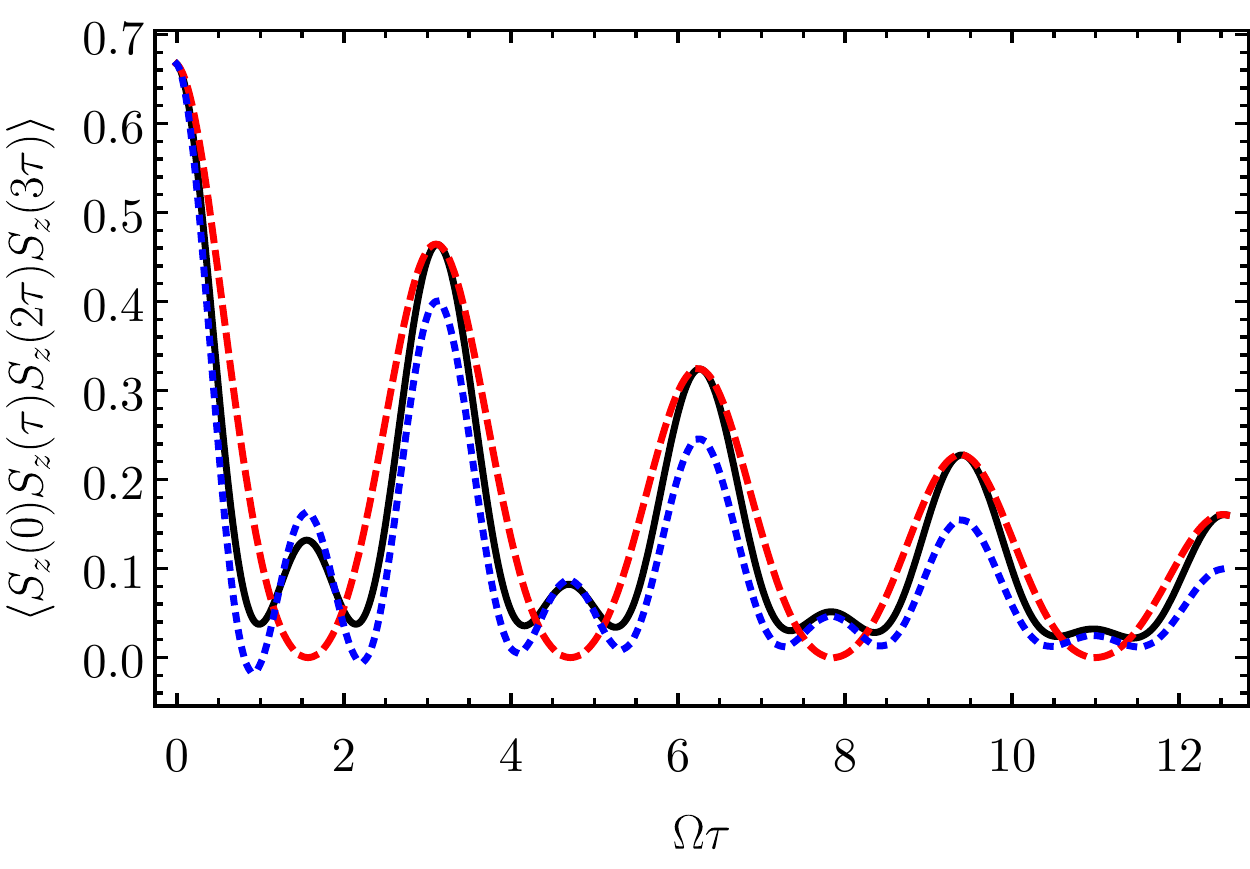}
  \caption{Fourth order correlator of a single spin $S=1$ for weak measurements [black solid curve, Eq.~\eqref{eq:s1_weak}], strong measurements [red dashed curve, Eq.~\eqref{eq:s1_strong}], and for a large ensemble of spins $S=1$ [blue dotted curve, Eq.~\eqref{eq:s1_norm}].}
  \label{fig:corr4}
\end{figure}

As a simple nontrivial example illustrating the difference between strong and weak measurements from the point of view of spin noise spectroscopy, one can consider the correlators of spin $S$ in transverse magnetic field. The second order correlator does not depend on the measurement strength $\lambda$ and can be calculated using the methods described in Sec.~\ref{sec:methods}. For $S=1/2$ the fourth order correlator does not depend on $\lambda$ as well. However, the difference shows up for $S=1$. For example, for three equal time intervals one can show that
\begin{subequations}
  \begin{multline}
    \label{eq:s1_weak}
    \braket{\delta S_z(0)\delta S_z(\tau)\delta S_z(2\tau)\delta S_z(3\tau)}=\frac{1}{36}\left[16\cos^2(\Omega_L \tau)\e^{-2\tau/\tau_s}+\e^{-3\tau/\tau_s}+\cos(2\Omega_L\tau)\e^{-3\tau/\tau_s}
    \right.\\\left.
      +6\cos(4\Omega_L\tau)\e^{-3\tau/\tau_s}\right],
  \end{multline}
  \begin{equation}
    \label{eq:s1_strong}
    \braket{\delta S_z(0)\delta S_z(\tau)\delta S_z(2\tau)\delta S_z(3\tau)}_{\rm strong}=\frac{1}{18}\cos^2(\Omega_L\tau)\e^{-2\tau/\tau_s}\left[8+\e^{-\tau/\tau_s}+3\cos(2\Omega_L\tau)\e^{-\tau/\tau_s}\right],
  \end{equation}
\end{subequations}
where the first expression describes the ``usual'' correlator~\eqref{eq:corr_n_sym}, which corresponds to the weak measurements, and the subscript <<$\rm strong$>> corresponds to the limit $\lambda\to\infty$. The spin relaxation here is described in the ``$\tau$-approximation'' for the density matrix. These correlators are shown in Fig.~\ref{fig:corr4} by the black solid and red dashed lines, respectively.

In the limit of many particles, the spin noise is Gaussian, so the fourth order cumulant vanishes and the fourth order correlator reduces to the sum of the products of the second order correlators, see Eq.~\eqref{eq:cum4}. In this case, we obtain (the spectrum is normalized by the the number of spins)
\begin{multline}
  \label{eq:s1_norm}
  \braket{\delta S_z(0)\delta S_z(\tau)\delta S_z(2\tau)\delta S_z(3\tau)}_{\rm Gauss}=\frac{1}{9}\left[2\cos^2(\Omega_L\tau)\e^{-2\tau/\tau_s}
        \right.\\\left.
          +\e^{-4\tau/\tau_s}+\cos(2\Omega_L\tau)\e^{-4\tau/\tau_s}+2\cos(4\Omega_L\tau)\e^{-4\tau/\tau_s}\right].
\end{multline}
This expression is shown by the blue dotted curve in Fig.~\ref{fig:corr4}.

In the case of the usual Faraday rotation measurement for a single spin, the parameter $\lambda$ can be estimated as $\overline{N}\theta_F^2$~\cite{Liu2010}, where $\overline{N}$ is the average number of photons in the probe pulse and $\theta_F$ is the Faraday rotation angle for completely polarized spin. For typical experiments, the authors of Ref.~\cite{Liu2010} give an estimate $\lambda\sim10^{-4}$. Measurement of the high order spin correlators allows one, for example, to distinguish between homogeneous and inhomogeneous broadening of the spin resonances (spin dephasing and decoherence times)~\cite{Liu2010,starosielec2010two,4order_exp}, study the properties of the reservoir, that leads to the spin relaxation~\cite{PhysRevB.90.205419,SCHAD2015401}, and study the effects of the interaction, that are not accessible from the second order correlator~\cite{Sinitsyn-Correlators}. For electrons localized in quantum dots, the measurement of the fourth order spin  correlators allows one to obtain the parameters of the nuclear spin dynamics caused by the precession in the external magnetic field, in the Knight field or by the interaction of the nuclear quadrupole moment with the elastic strain in the quantum dot~\cite{PhysRevB.96.045441,PhysRevB.99.155305}.

In the case of monoexponential spin relaxation, the usual spin noise spectrum consists of the series of the Lorentzian peaks with the corresponding widths centered at the eigenfrequencies of the system, see Eq.~\eqref{eq:spectrum_num}. Interestingly, the decomposition of the free energy into powers of spin operator allows one to show that the spectrum of the correlators of high orders is also described by universal expressions with a small set of parameters~\cite{SNS_Universality}. The same approach allows one to use the time reversal symmetry to establish the general relations between the spin correlation functions of high orders and nonlinear spin susceptibility or dependence of the lower order spin correlators on magnetic field~\cite{Bochkov,sinitsynreview}.  

Typically, the spin noise is not measured using the probe pulses, but using the continuous light. In this case, the spin measurements at intermediate times do not directly contribute to the correlator~\eqref{eq:corr_Krauss}, but modify the density matrix~\cite{Liu2010}. This was demonstrated in the measurement of the second order spin correlation function for a single quantum dot~\cite{4order_exp}. Thus, for example, if the strong spin measurement is performed at the time moment $\tau$ between the two other spin measurements at times $0$ and $t$ ($0<\tau<t$), then the correlator takes the form
\begin{equation}
  \braket{S_z(0)S_z(t)}_\tau=4\braket{S_z(0)S_z^2(\tau)S_z(t)},
\end{equation}
where the spin $1/2$ is considered. For the strong intermediate measurement, this correlator is generally different from $\braket{S_z(0)S_z(t)}$~\cite{PhysRevB.96.045441,PhysRevB.99.155305}.

In the limit of strong continuous measurements, the spin dynamics is almost frozen due to the quantum Zeno effect, as it was described in Sec.~\ref{sec:4level}. The spin noise in this conditions is the telegraph noise: at each moment the spin is in one of the eigenstates~\cite{Zeno_QED}.

In conclusion, we note that the direct measurement of the high order spin correlators is challenging, and a number of alternative detection methods are suggested. For example, to overcome the parametric suppression of the high order cumulants for many spins, one can study the ``stimulated'' spin noise~\cite{doi:10.1063/1.5116901}, when external magnetic field~\cite{PhysRevA.93.033814} or optical orientation synchronizes single spins, so that they are no longer independent. The other possibility of high order spin noise measurement is due to the nonlinear relation between the detected  Faraday rotation angle and the spin polarization. This method allowed the authors of Ref.~\cite{fomin2019spinalignment} to observe the noise of optical alignment in atomic gases. The same method allows one to study the spin correlators of high orders for the spins, that do not directly participate in the optical transitions~\cite{smirnov2020optical}. Such spins can be provided by the host lattice nuclei, magnetic impurities, nuclei of donors and acceptors, which created electron or hole bound states~\cite{artemova85,Nuclear_Faraday}.

In typical experimental conditions, the dominant mechanism of the Faraday rotation by nuclear spin fluctuation $\delta \bm I$ is related with the splitting of the trion resonance frequency $\omega_0$ by $2a\delta I_z$ due to the hyperfine interaction~\cite{NuclearNoise,smirnov2020optical} ($a$ is the real constant). For $\sigma^\pm$ polarized light at frequency $\omega$ the contribution to the transmission coefficient caused by a single resonance has the form
\begin{equation}
  \label{eq:tpm}
  t_\pm\propto\frac{1}{\omega-\omega_0\mp a\delta I_z+\i\gamma},
\end{equation}
where $\gamma$ is the homogeneous resonance linewidth. The Faraday rotation angle is determined by the difference of the phases of the transmitted circularly polarized components and has the form
\begin{equation}
  \label{eq:theta_nucl}
  \theta_F\propto\Im\left[(t_+-t_-)t^*\right]\propto\frac{a\delta I_z\gamma}{(\omega-\omega_0)^2+\gamma^2},
\end{equation}
where $t=(t_++t_-)/2$, and we took into account the fact that $a\delta I_z\ll\gamma$. One can see, that in contrast with the Faraday rotation by spin noise of the resident charge carriers~\eqref{dtheta:res} this effect is the strongest exactly at the resonance ($\omega=\omega_0$). To describe the Faraday rotation in realistic systems, Eq.~\eqref{eq:theta_nucl} can be averaged over the inhomogeneous broadening of the resonance. Alternative mechanisms of the Faraday rotation are the variations of the thermal occupancies of the electron spin sublevels and contributions from the interband transitions~\cite{artemova85,Nuclear_Faraday}.

For example, at zero detuning, $\omega=\omega_0$, the Faraday rotation signal [see Eqs.~\eqref{dtheta:res} and~\eqref{eq:theta_nucl}] can be presented as
\begin{equation}
  \mathcal F=\Re\{ \delta P_y E_{0,x}^*\}=C\sum_{n=0}^\infty\left(\frac{a\delta I_z}{\gamma}\right)^{2n+1},
\end{equation}
where $C$ is a real constant. This expressions shows that even the second order correlation function of the Faraday rotation contains contributions from the spin correlation functions of high orders:
\begin{equation}
  \braket{\mathcal F(0)\mathcal F(\tau)}=C^2\sum_{n,n'}\left(\frac{a}{\gamma}\right)^{2(n+n'+1)}\braket{{\delta I_z^{2n+1}(0)\delta I_z^{2n'+1}(\tau)}}.
\end{equation}
The drawback of this approach is the possibility to measure the spin correlators related with the two time moments $0$ and $\tau$ only.




\section{Extensions of spin noise spectroscopy}
\label{sec:extended}

Most often, the spin noise spectroscopy, as described above, is used to study the spin properties of electrons and holes. In the same time, the Faraday rotation can be induced by any physical quantity, which transforms under symmetry operations in the same way as the component of the pseudovector $S_z$. In this section we describe the abilities to detect the fluctuations of the host lattice nuclear spins, magnetic impurities, electric charges in magnetic field, valley polarization and electric current in gyrotropic systems. Apart from that, we discuss the opportunities to study optical spectra and spatial correlations of spin noise, which were realized experimentally in the past few years.

\begin{figure}
  \centering
  \includegraphics[width=0.3\linewidth]{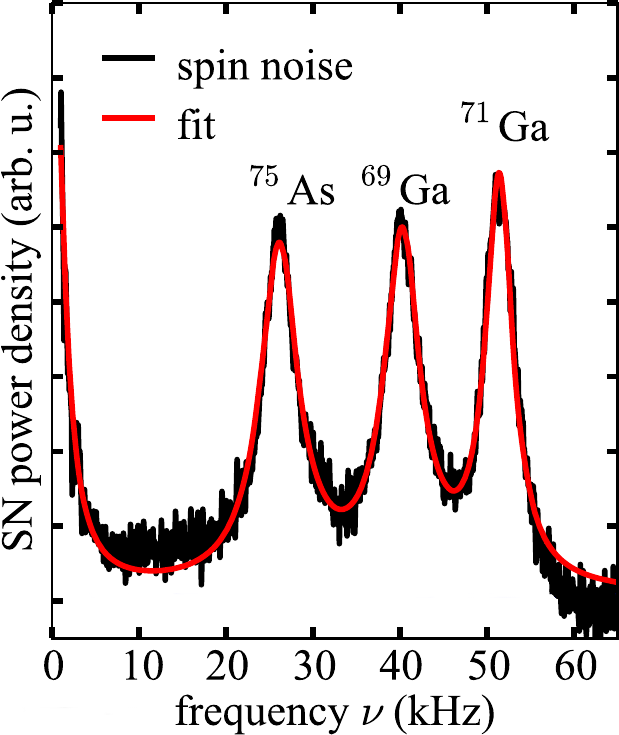}
  \caption{Nuclear spin noise spectrum detected in the vicinity of the trion resonance for electrons bound at Si donors in bulk GaAs in transverse magnetic field of $3.75$~mT.  (Adapted from Ref.~[\onlinecite{NuclearNoise}].)}
  \label{fig:nuclei_noise}
\end{figure}

Spins of the host lattice nuclei do not directly participate in the interband optical transitions. However, the mechanism of the Faraday rotation for nuclei discussed in the previous section allows one to detect their spin noise. For example, in Fig.~\ref{fig:nuclei_noise} we show the Faraday rotation noise spectrum for bulk GaAs doped with Si for the frequency of the probe beam in the vicinity of the trion resonance for donors. One can distinctly see the peaks at the spin precession frequencies of nuclei of $^{69}$Ga, $^{71}$Ga, and $^{75}$As~\cite{NuclearNoise}. In this system, the hyperfine interaction with the resident electron and hole in trion leads to the shift of the trion resonance by
\begin{equation}
  a\delta I_z=\frac{1}{\hbar}\left(\frac{1}{2}A^e+\frac{3}{2}A^h\right)\sum_{k=1}^{N_n}\delta I_{k,z},
\end{equation}
as discussed in the previous section. Here the hyperfine interaction constants for electron, $A^e$, and hole in trion, $A^h$, are assumed to be equal for all  nuclei in the vicinity of the donor, cf. Eq.~\eqref{eq:Omega_N}. Similarly, the hyperfine interaction allows one to detect the spin noise of magnetic impurities, for example, manganese~\cite{Scalbert2015,smirnov2020optical}.

Nuclear spin noise spectra can be calculated, for example, in the framework of the central spin model (see Sec.~\ref{sec:central}). The spin noise spectrum of nuclei with $I=1/2$ in the strong transverse magnetic field has the form~\cite{PhysRevB.97.195311}
\begin{equation}
  \braket{{\delta} I_z^2}_\omega=\frac{\pi\hbar N_n}{2}\mathcal P(|2\hbar\omega-\mu_ng_nB|),
\end{equation}
where $\mathcal P(A)$ is the distribution function of the hyperfine interaction constants. Thus, the shape of the spin noise spectrum, in contrast with the electron spin noise spectrum, allows one to determine the distribution function not of the Overhauser field, but of the Knight field.

This approach allows one to study the nuclear spin dynamics at the submillisecond time scale. However, the nuclear spin dynamics can take place at the time scale of the order of minutes, or even hours in the case of dynamic nuclear polarization or relaxation of nuclear polarization. Since the rate of the energy transfer between the nuclear spins is much faster than that between nuclei and the host lattice or resident electrons,  the nuclear spin system can be described using the effective nuclear spin temperature $\Theta_N(t)$, which slowly varies with time. In realistic systems this temperature can be smaller than the temperature of the host lattice by a few orders of magnitude and be positive or negative~\cite{abragam_rus}.

To study experimentally the nuclear spin dynamics, in the first stage the nuclei should be dynamically polarized in the longitudinal magnetic field using the strong circularly polarized beam, which induces interband transitions in the semiconductor. In the second stage, the optical excitation should be switched off and magnetic field can be optionally reoriented. Nuclear spin temperature $\Theta_N(t)$ slowly relaxes with the laboratory time $t$ to the lattice temperature, and the electron spin noise of resident electrons is measured during this time. In transverse magnetic field (Voight geometry) the precession peak in the spin noise spectrum is centered at the frequency $\bm\Omega_{\rm tot}(t)=\bm\Omega_B+\overline{\bm\Omega}_N(t)$ [cf. Eq.~\eqref{eq:transverse}]. Here
\begin{equation}
  \overline{\bm\Omega}_N(t)=\frac{AI}{\hbar}\frac{\bm B}{B}\mathcal B_I\left(\frac{\mu_n g_n B I}{k_B\Theta_N(t)}\right)
\end{equation}
is the spin precession frequency in the Overhauser field average over the ensemble~\cite{glazov2018electron}, where $A$ is the hyperfine interaction constant, $I$ is the nuclear spin, $\mathcal B_I(x)$ is the Brillouin function, $\mu_n$ and $g_n$ are the nuclear magneton and $g$-factor, respectively.

\begin{figure}
  \centering
  {\includegraphics[width=0.6\textwidth]{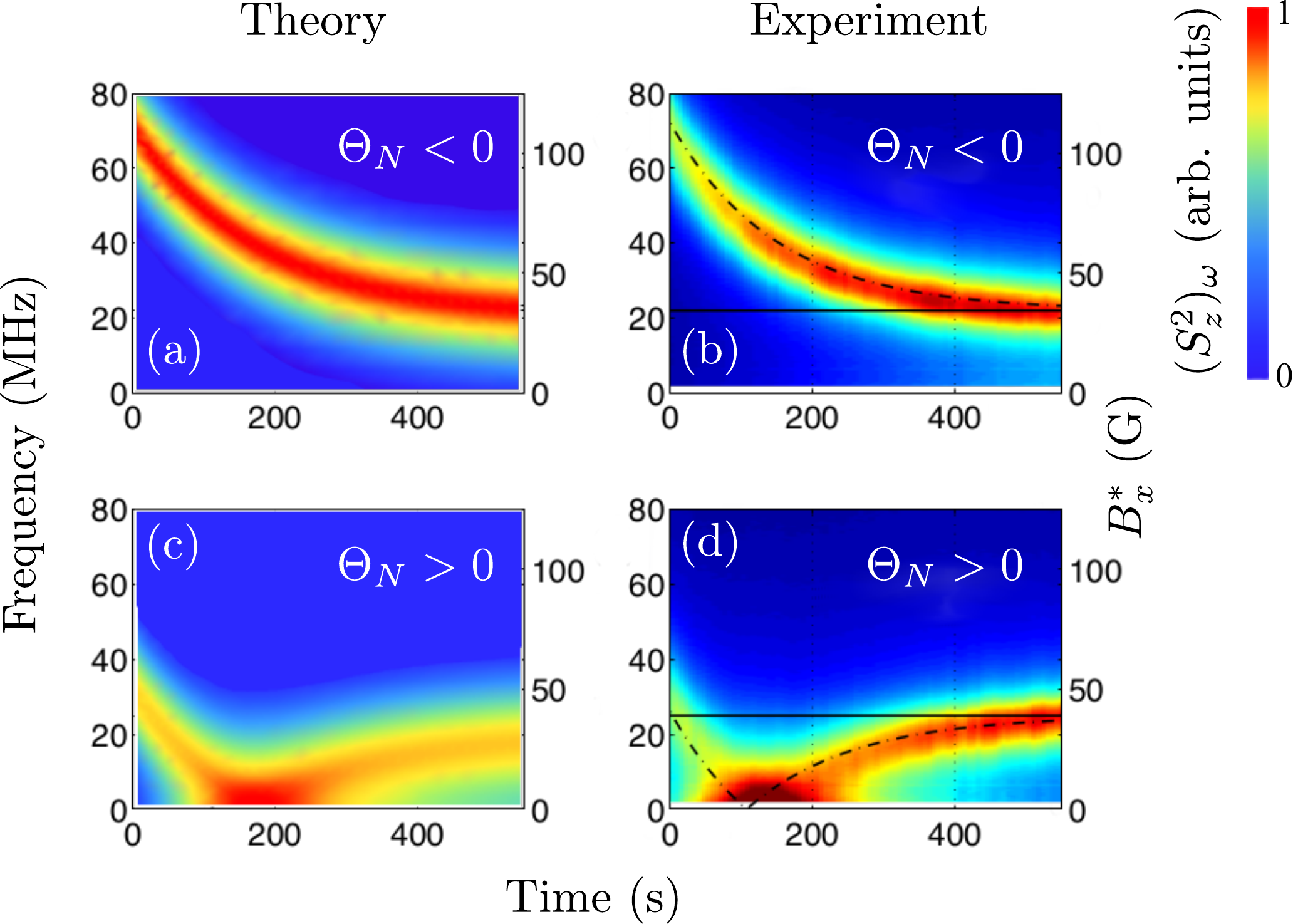}}
  \caption{Time resolved spin noise spectra of electrons, interacting with the nonequilibrium nuclear spin polarization perpendicular to the probe beam propagation direction. Panels (a), (b) and (c), (d) correspond to the positive and negative nuclear spin temperatures, respectively. Panels (a), (c) and (b), (d) show theoretical and experimental results, respectively. (Adapted from Ref.~[\onlinecite{NuclearPolarization}].)}
  \label{fig:nuclei}
\end{figure}

Since the nuclear spin temperature $\Theta_N(t)$ relaxes to the host lattice temperature, the measurement of the electron spin noise spectrum with the time resolution allows one to study nonequilibrium nuclear spin polarization in the absence of external excitation~\cite{polarizednuclei}. This proposal was realized experimentally~\cite{NuclearPolarization,OpticalField}. The theoretical and experimental results are shown in Fig.~\ref{fig:nuclei} in panels (a,c) and (b,d), respectively. The color shows the intensity of the nuclear spin noise at the given frequency at the given time moment. Depending on the sign of the nuclear spin temperature, the spin precession frequency $\Omega_{\rm tot}(t)$ can be either always positive [panels (a) and (c)], or cross zero [panels (b) and (d)]. Interestingly, the electron excitation by circularly polarized light creates for them an effective longitudinal magnetic field due to the dynamic Zeeman effect~\cite{OpticalField}. The width of the precession peak, as mentioned in Sec.~\ref{sec:0D}, generally depends on the nuclear polarization degree~\cite{polarizednuclei}.

\begin{figure}
  \centering
  \includegraphics[width=0.7\linewidth]{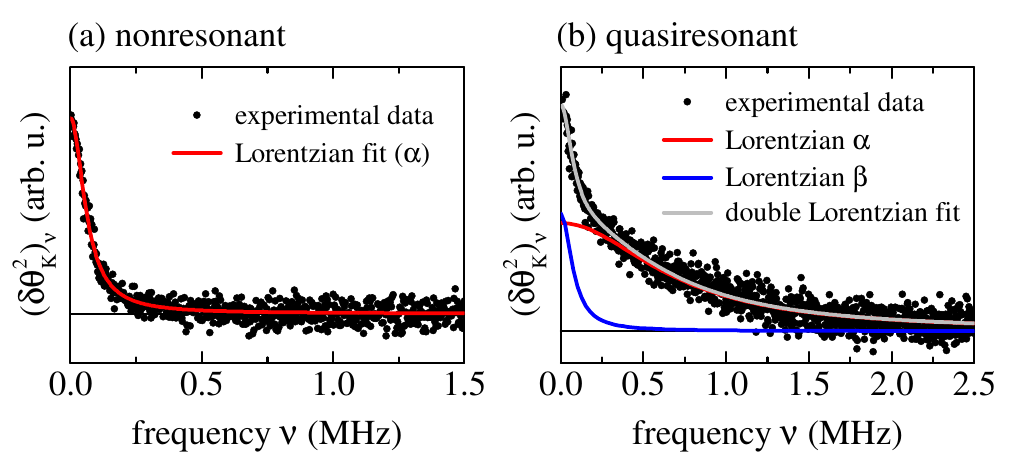}
  \caption{Kerr rotation noise spectra for (a) $|\omega-\omega_0|\gg\gamma$ and (b) $|\omega-\omega_0|\ll\gamma$ in external longitudinal magnetic field $B_z=31$~mT, measured for a single In(Ga)As quantum dot. The blue curve shows the contribution from the quantum dot occupancy noise. (Adapted from Ref.~\cite{Occupancy-noise}.)}
  \label{fig:charge_noise}
\end{figure}

Similarly to the nuclear spin polarization, an external magnetic field $\bm B$ also leads to the Faraday rotation described by Eq.~\eqref{eq:theta_nucl}, where the role of the splitting $2a\delta I_z$ is played by $(g_e-g_h)\mu_BB_z/\hbar$ ($g_e$ and $g_h$ are the g-factors of the electron and hole in trion). This Faraday rotation takes place only if the localized state is occupied with a single charge carrier. If the occupancy of the state can take two values $n=0,1$, its fluctuations $\delta n$ lead to the noise of the Faraday signal:
\begin{equation}
  \label{eq:charge_noise}
  \delta\theta_F\propto\frac{B_z\gamma}{(\omega-\omega_0)^2+\gamma^2}\delta n.
\end{equation}
In Fig.~\ref{fig:charge_noise} we show the spectra of Kerr rotation measured for a single quantum dot~\cite{Occupancy-noise,ChargeNoise}. Far from the resonance [panel~(a)] the spectrum consists of the peak at zero frequency, which corresponds to the hole spin noise. With approach of the resonance [panel~(b)] this peak broadens [see Sec.~\ref{sec:4level}] and in agreement with Eq.~\eqref{eq:charge_noise} the new peak appears, which is related with the fluctuations of the quantum dot occupancy (blue curve). The detailed analysis of the spectra allows one to determine the Auger recombination rate of the trion in addition to the parameters of the spin dynamics.

Additional extensions of the spin noise spectroscopy are provided by the spin-orbit interaction, which can lead to the lifting of the spin degeneracy in the many-valley semiconductors. In this case, the time reversal symmetry changes the Bloch wave vector to the opposite one, so all the states remain two fold degenerate in agreement with the Kramers theorem. A pair of the degenerate states can be characterized by the valley pseudospin $\bm \tau$. Thus, in many-valley semiconductors, Faraday rotation noise spectrum consists of two contributions~\cite{Valley_Noise}
\begin{equation}
  (\delta\theta_F^2)_\Omega=A(\delta S_z^2)_\omega+B(\delta\tau_z^2)_\omega,
\end{equation}
which describe the spin and valley noise, respectively (there can also be a contribution from the cross-correlation functions $\braket{\delta S_z(t)\delta\tau_z(t')}$). For the transition metal dichalcogenide monolayers, the microscopic theory of this effect was developed~\cite{Valley_Noise}, and the valley noise was measured~\cite{Gorycaeaau4899}.

The spin-orbit interaction can also lead to the locking of the spin direction with the direction of the electron propagation. The classical example is bulk tellurium, where the current flow leads to the spin polarization and induces the optical activity~\cite{ivchenko1978new,Baranova1977243,vorob1979optical}:
\begin{equation}
  \theta\propto j_z\propto S_z,
\end{equation}
where the $z$ axis is the main axis of the crystal and also the light propagation direction. In this case the Faraday rotation noise spectrum is determined by the current noise spectrum~\cite{GyrotropySmirnov}. The effects of this kind are possible in any gyrotropic system, for example, for ensembles of chiral nanotubes~\cite{ivch_spi}, and GaAs-based quantum wells with the crystallographic orientation $[001]$ for the oblique incidence of light~\cite{Kotova2016} (in this case the contribution of the spin current noise is also possible). An illustrative example is given by the quantum wells with the $[110]$ crystallographic orientation with the $C_{2v}$ point symmetry group, where the spin-orbit interaction Hamiltonian allows for the contributions of the form~\cite{dyakonov:110,PhysRevB.85.205307}
\begin{equation}
  \mathcal H_{SO}=\beta_e\sigma_z^{(e)}k_x^e+\beta_h\sigma_z^{(h)}k_x^h.
\end{equation}
Here $\beta_{e,h}$ are the parameters of the spin-orbit interaction for the lowest electron and hole subbands, $\sigma_{z}^{(e,h)}$ are the Pauli matrices acting of the pseudospin, and $k_{x}^{e,h}$ are the components of the corresponding wave vectors ($x\parallel[\bar{1}10]$, $y\parallel[001]$). From this expression one can see, that $k_{x}$ transforms in the same was as $\sigma_z$, so the fluctuations of the current along the $x$ axis can lead to the rotation of the polarization plane of the incident light along with the spin noise. After the reflection of the incident wave, the polarization conversion is determined by the off-diagonal reflection coefficient~\cite{GyrotropySmirnov}
\begin{equation}
  r_{xy}=-r_{yx}=P_s s_z+P_j j_x,
\end{equation}
where $\bm s$ and $\bm j$ are the spin and current densities, respectively, and the coefficients $P_s$ and $P_j$ in the spectral range around the transition between the lowest electron and hole subbands are
\begin{equation}
  P_s=\frac{2\pi\omega|d|^2}{c(E_0-\hbar\omega)}, \quad P_j=\frac{\pi\omega|d|^2\beta m_c}{ec\hbar(E_0-\hbar\omega)^2}.
\end{equation}
Here $d$ is the transition dipole moment, $E_0$ is the transition energy (the damping is neglected), $\beta=\beta_e-\beta_h$, $m_c$ is the electron effective mass, and the valence band is assumed to be completely occupied. Thus, the Kerr rotation noise spectrum consists of the contributions from the spin noise, current noise, and cross-correlations. In this case, in the vicinity of the resonance, the dominant contribution is related with the current noise and far from the resonance with the spin noise.

The measurement of the Faraday rotation noise intensity as a function of the detection frequency $\omega$ is termed optical spin noise spectroscopy~\cite{Zapasskii13}. This method allows one to separate the different contributions to the noise of optical signals and to distinguish between homogeneous and inhomogeneous broadenings of the optical resonances. Indeed, in the case of the homogeneous broadening, the noise intensity is proportional to the squared Faraday rotation:
\begin{equation}
  \label{eq:noise_power}
  \braket{\delta\theta_F^2}\propto\frac{(\omega-\omega_0)^2}{(\omega-\omega_0)^2+\gamma^2}\braket{\delta S_z^2},
\end{equation}
see Eq.~\eqref{dtheta:res}, and has a dip at the resonance frequency at $\omega=\omega_0$. If the inhomogeneous broadening with the typical width $\Delta\omega_0$ is present in the system, then this expression should be averaged with the corresponding distribution. For example, for the Gaussian broadening with $\Delta\omega_0\gg\gamma$ we obtain
\begin{equation}
  \braket{\delta\theta_F^2}\propto\exp\left(-\frac{(\omega-\omega_0)^2}{\Delta\omega_0^2}\right)\braket{\delta S_z^2},
\end{equation}
so the noise intensity is the largest at the center of the inhomogeneously broadened line. Interestingly, in atomic vapours, where the inhomogeneous broadening is provided by the Doppler effect, there can be a dip at the center of the inhomogeneously broadened line similarly to the case of homogeneous broadening, which is caused by the fast momentum relaxation of atoms~\cite{PhysRevA.97.032502}.

Noteworthy, the proportionality between the spin noise intensity and the number of the charge carriers in the probed volume [see, for example, Eq.~\eqref{corr:degen}] allows one to determine the distribution of the concentration of the charge carriers by moving the focus of the laser beam~\cite{Z-scan}.

The two probe beams instead of one allow one to analyze the cross-correlations of the Faraday rotation for the two beams~\cite{SpinNoiseCollisions,roy2015cross}. These fluctuations are the largest, when the two beams are crossed~\cite{Zapasskii:13,Kozlov:18}, but even if their intersection is absent, the cross-correlations can take place due to the ballistic or diffusive propagation of electrons from one place to another~\cite{Sinitsyn_2Beam,poshakinskiynoise}, see Sec.~\ref{sec:2D}. The measurements of this kind allow one to study the spin dynamics not only with the time resolution, but also with the spatial resolution, and to determine, for example, parameters of the spin-orbit interaction for electrons and holes in the quantum wells.

\begin{figure}
  \centering
  \includegraphics[width=0.6\linewidth]{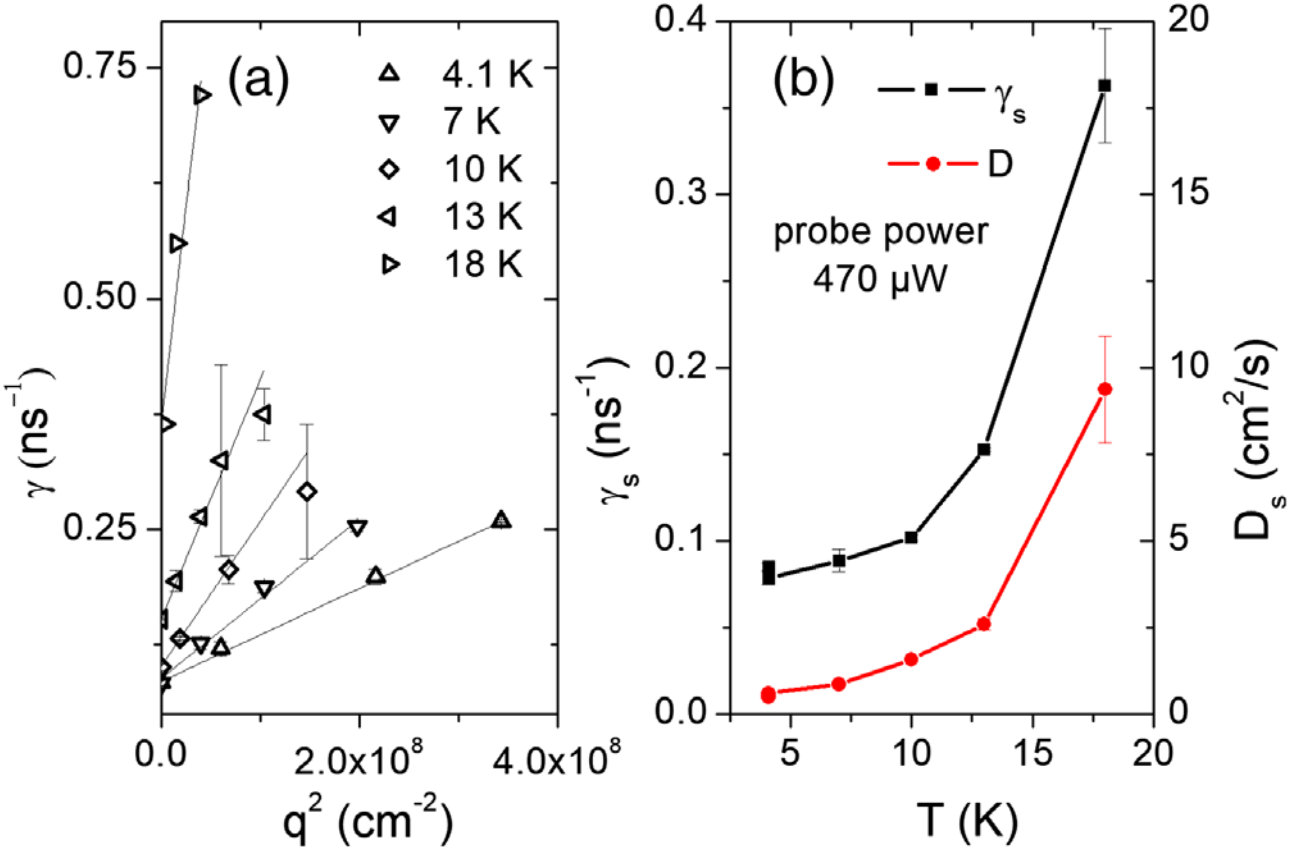}
  \caption{Width of the spin noise spectrum $(\delta S_z^2)_{\omega,q}$ as a function of (a) $q$ and (b) temperature, measured for the bulk $n$-type CdTe. The red dots in panel (b) show the electron spin diffusion coefficient, determined from these measurements. (Adapted from Ref.~\cite{PhysRevLett.123.017401}.)}
  \label{fig:Ds}
\end{figure}

Presently, a few experimental groups are studying the possibility of the spatial resolution of spin noise based on the close relation of the spin noise spectroscopy with the Raman spin-flip scattering, see Sec.~\ref{sec:Raman}. On this way, the interference is studied between the probe light after propagation through the sample and the additional (reference) beam, which homo- or heterodynes the signal~\cite{PhysRevA.95.043810,PhysRevLett.123.017401}. In this case, the Stokes parameter that is proportional to the Faraday rotation angle [see Eq.~\eqref{dtheta:res}], has the form
\begin{equation}
  \delta\xi_1(t)=\frac{2\Re[\delta E_y^{pr}(t)E_{0,x}^*]}{|E_{0,x}^2|},
\end{equation}
where the probe and reference beams are polarized along the $x$ axis, $E_{0,x}$ is the amplitude of the reference beam, and $\bm E^{pr}$ is the amplitude of the beam transmitted through the sample, and it is assumed that $E_{0,x}\gg E^{pr}$. Experimentally, this method can be used to increase the sensitivity of the spin noise measurement~\cite{PhysRevB.97.125202} and to analyze the high-frequency spin noise~\cite{:/content/aip/journal/rsi/87/9/10.1063/1.4962863}. Importantly, after the scattering of the probe beam with the change of the wave vector by $\bm q$ the Faraday rotation is determined not by the total spin fluctuation, but by its spatial harmonic $\delta S_z(t,\bm q)$~\cite{PhysRevLett.123.017401}:
\begin{equation}
  \delta\xi_1(t)\propto\int s_z(t,\bm r)\e^{-\i\bm{qr}}\d{\bm r}\equiv\delta S_z(t,\bm q).
\end{equation}
For example, provided the spin dynamics is characterized by the spin relaxation time $\tau_s$ and the diffusion coefficient $D_s$, the Faraday rotation noise spectrum has the usual form
\begin{equation}
  (\delta S_z^2)_{\omega,q}=\frac{\tau/2}{1+(\omega\tau)^2}
\end{equation}
[cf.~\eqref{eq:longitudinal}], however its width depends on the wave vector:
\begin{equation}
  \frac{1}{\tau}=\frac{1}{\tau_s}+D_sq^2.
\end{equation}
The measurements of this kind allow one to study the transition from localized to free electrons with increase of the temperature by the change of the diffusion coefficient~\cite{cronenberger2019spatiotemporal}. An example, of the experimental measurement of the width of the spin noise spectrum as a function of the wave vectors and of the temperature dependence of the diffusion coefficient is shown in Fig.~~\ref{fig:Ds} for the case of electrons localized at donors in the bulk cadmium telluride.





\section{Conclusion and outlook}
\label{sec:concl}


Theory of spin noise in low dimensional systems and bulk semiconductors is reviewed. Spin noise in such systems is usually detected by the fluctuations of Faraday rotation of the continuous probe beam. General theoretical approaches were illustrated by a number of experimental results for the structures of various dimensionality from 0D to 3D. At the same time, the review contains a number of original results concerning the influence of electrons tunneling on the spin noise spectra and many-body spin localization (section~\ref{sec:tun}), spin fluctuations in 1D systems (section~\ref{sec:1D}) and calculation of fourth order spin correlators depending on the strength of measurements (section~\ref{sec:high}).

Further perspectives of the development of the spin noise spectroscopy method follow from the sections ~\ref{sec:high} and~\ref{sec:extended}. Firstly, they are related to the analysis of the high order spin correlations, secondly, with the development of the experimental technique for the spin noise measurement with the spatial resolution, and thirdly, measurement of the charge and valley fluctuations of charge carriers in semiconductors. Moreover, as the sensitivity in the experiments can be increased by the homo- and heterodyne detection~\cite{:/content/aip/journal/rsi/87/9/10.1063/1.4962863,PhysRevB.97.125202} and by using the squeezed light~\cite{PhysRevA.93.053802}, it seems promising to combine these two experimental methods. Optomechanical resonances can be also applied for the increase of noise signal, as it was previously done in  magnetometry~\cite{Li:enhanced}.

The manifestation of quantum back action during the spin noise measurements is still not well understood. In spite of it, the quantum Zeno effect, probably, has already been observed experimentally~\cite{noise-trions}, its microscopic description and the link between the strength of measurements and measured spin signal is still not clear~\cite{Zeno_QED}. A fundamental challenge for the theory is the description of spin fluctuations in mesoscopic systems. Nowadays, the existing models can describe only spin noise spectra for a small number of spins or for a large ensemble of spins. In the mesoscopic case, for example, when the spin of an electron interacts with several dozens of nuclei or magnetic impurities, the theory remains undeveloped, and theoretical predictions of new effects in such systems are not yet available. Another problem deals with the description of spin noise in the nonlinear systems, for example, in the vicinity of phase transition or at formation of spin polarons. Nonlinear equations of motion can also describe the nuclear field that acts on the localized electrons, as it is a classical quantity, not quantum one.

The evolution of the theory of spin fluctuations is mostly determined by the technological progress and the appearance of new systems in which new unexpected effects are predicted and observed. From this point of view, one of the most promising systems for further investigation of spin noise are twisted Van-der-Waals structures based on the monolayers of transition metals dichalcogenides, as well as perovskites, topological insulators, Weyl semi-metals and organic semiconductors.

Authors are thankful to V. S. Zapasskii and G. G. Kozlov for fruitful discussions and careful reading of the review. The reported study was funded by RFBR, project number 19-12-50293.


\renewcommand{\i}{\ifr}

\bibliography{bib_eng.bib}

\end{document}